\RequirePackage{rotating}
\documentclass[structabstract]{aa}
\pdfoutput=1

\usepackage{savesym}
\usepackage{caption}
\usepackage{amsmath}
\usepackage{xcolor}
\savesymbol{iint}
\usepackage{txfonts}
\restoresymbol{TXF}{iint}
\usepackage{natbib}
\usepackage{graphicx}
\usepackage{rotating}
\usepackage{tablefootnote}
\usepackage{threeparttable}
\usepackage{aalongtable}
\usepackage{hyperref}
\usepackage{float}
\usepackage{placeins}
\usepackage{array}
\newcommand{\Msun}{$M_{\odot}$}

\newcommand{\Lsun}{$L_{\odot}$}

\newcommand{\Mstar}{$M_{\star}$\ }
\newcommand{\Lstar}{$L_{\star}$\ }
\newcommand{\Rstar}{$R_{\star}$\ }

\newcommand{\Rsun}{$R_{\odot}$}

\newcommand{\Ox}{[\oi]$\lambda$6300}

\newcommand{\Ha}{H$\alpha$}
\newcommand{\Hb}{H$\beta$}

\newcommand{\sii}{\ion{S}{ii}}
\newcommand{\oi}{\ion{O}{i}}
\newcommand{\oii}{\ion{O}{ii}}
\newcommand{\oiii}{\ion{O}{iii}}

\newcommand{\nii}{\ion{N}{ii}}

\newcommand{\hei}{\ion{He}{i}}

\newcommand{\neiii}{\ion{Ne}{iii}}

\newcommand{\feii}{\ion{Fe}{ii}}
\newcommand{\caii}{\ion{Ca}{ii}}

\newcommand{\SII}{[\sii]$\lambda\lambda$6716,6731}
\newcommand{\NII}{[\nii]$\lambda$6583}
\newcommand{\OIII}{[\oiii]$\lambda$5007}
\newcommand{\OI}{[\oi]$\lambda\lambda$6300,6363}
\newcommand{\Feiia}{[\feii]$\lambda$7155}
\newcommand{\Feiib}{[\feii]$\lambda$8617}
\newcommand{\OIa}{[\oi]$\lambda$6300}

\newcommand{\niia}{[\nii]$\lambda$6583}
\newcommand{\siia}{[\sii]$\lambda$6716}
\newcommand{\siib}{[\sii]$\lambda$6731}
\newcommand{\siirat}{[\sii]$\lambda\lambda$6716/6731}

\newcommand{\mh}{H$_{2}$\ }

\newcommand{\kms}{km s$^{-1}$}
\newcommand{\ndeg}{$^{\circ}$}

\newcommand{\av}{A$_{V}$}
\newcommand{\eden}{$n_{\mathrm{e}}$}
\newcommand{\xe}{$x_{\mathrm{e}}$}
\newcommand{\Te}{$T_{\mathrm{e}}$}
\newcommand{\Lacc}{$L_{\mathrm{acc}}$}
\newcommand{\Macc}{$\dot{M}_{\mathrm{acc}}$}

\newcommand{\Mout}{$\dot{M}_{\mathrm{out}}$}
\newcommand{\Moutnh}{$\dot{M}_{\mathrm{out,nH}}$}
\newcommand{\MoutLo}{$\dot{M}_{\mathrm{out,L[\oi]}}$}
\newcommand{\MoutLsii}{$\dot{M}_{\mathrm{out,L[\sii]}}$}
\newcommand{\Moutshock}{$\dot{M}_{\mathrm{out,shock}}$}

\newcommand{\peryr}{yr$^{-1}$}
\newcommand{\nh}{$n_{\mathrm{H}}$}

\newcommand{\percm}{cm$^{-3}$}
\newcommand{\vrad}{$v_{\mathrm{rad}}$}
\newcommand{\vtan}{$v_{\mathrm{tan}}$}
\newcommand{\vjet}{$v_{\mathrm{j}}$}
\newcommand{\vjetb}{$v_{\mathrm{j,b}}$}
\newcommand{\vjetr}{$v_{\mathrm{j,r}}$}
\newcommand{\vshock}{$v_{\mathrm{shock}}$}

\newcommand{\bunit}{$\times$ 10$^{-20}$ ergs s$^{-1}$ cm$^{-2}$}
\newcommand{\ergscm}{$\times$ ergs s$^{-1}$ cm$^{-2}$}

\usepackage{natbib}
\bibpunct{(}{)}{;}{a}{}{,}

\graphicspath{{./}}
\usepackage{graphicx,caption}
\usepackage{makecell}

\begin{document}
	\journalname{Astronomy $\&$ Astrophysics}
	\title{The accretion-ejection connection in the asymmetric Th 28 jet revealed by MUSE-NFM \thanks{Based on observations collected with MUSE at the Very Large Telescope on Cerro Paranal (Chile), operated by the European Southern Observatory (ESO). Program ID: 110.23ZG} }
	
	\author{
		A. \,Murphy \inst{1}
		\and 
		E. T. \,Whelan \inst{2}
		\and 
		F. \,Bacciotti \inst{3}
	    \and 
	 	A. \,Kirwan \inst{4}
        \and 
		D. \,Coffey \inst{4}
        \and 
		M. \,Birney \inst{5}
	 	\and 
	 	J. \,Eisl\"offel \inst{6}
        \and 
	 	M. \,Takami \inst{1}
        }
\institute{Academia Sinica Institute of Astronomy and Astrophysics, No 1. Sec. 4, Roosevelt Rd., Taipei 106319, Taiwan
\and
Department of Experimental Physics, Maynooth University, Maynooth, Co Kildare, Ireland
\and
INAF—Osservatorio Astrofisico di Arcetri, Largo E. Fermi 5, I-50125 Firenze, Italy
\and
 School of Physics, University College Dublin, Belfield, Dublin 4, Ireland
 \and
 ESO, Karl-Schwarzschild-Strasse 2, 85748 Garching bei M\"unchen, Germany
 \and
 Th\"uringer Landessternwarte, Sternwarte 5, 07778 Tautenburg, Germany
}

\titlerunning{MUSE-NFM Observations of the Th 28 jet} 
\date{Received 24 December 2025 / Accepted 23 January 2026}

\abstract {Mass loss through stellar jets is closely tied to the process of accretion through the disk. Understanding mass loss phenomena such as episodic ejections and outflow asymmetries can thus shed light on the mechanism of jet launching and its connection to both mass accretion and the evolution of the protoplanetary disk.} {We aim to determine the mass outflow rates close to the base of the asymmetric jets launched by the Classical T Tauri Star Th 28, and identify signatures of variable mass accretion/ejection in this source.} { We use new VLT/MUSE Narrow Field Mode observations of Th 28 to map the jet structures within 6\arcsec\ of the source at an effective angular resolution of 0\farcs12, provided by the combination of the AO correction and image deconvolution. The emission line profiles and flux ratios are investigated and diagnostic analysis of the optical forbidden emission lines (FELs) is used to estimate the electron density, ionisation fraction, electron temperature and shock velocities in both jet lobes within 200 au of the star. The mass outflow rates in each lobe are obtained using the derived total densities and FEL luminosities and compared with the mass accretion rate.} { We identify several new knots in both jet lobes which have been ejected in the previous 10 years. These indicate ejections on a timescale of 3-6 years, which is significantly more frequent than previously estimated. In both lobes we find comparable mass outflow rates close to the jet base, with average values of \Mout\ = 1-3 $\times$ 10$^{-8}$ \Msun\ \peryr. We find that the mass accretion rate has approximately doubled between 2014 and 2023, and estimate \Macc\ = 2.11 $\times$ 10$^{-7}$ \Msun\ \peryr. Analysis of the red-shifted jet mass outflow rate shows a similar increase of a factor 2, indicating that the ratio of mass outflow to accretion remains constant.} { Th 28 has undergone a significant rise in mass accretion rate since the previous epoch, which may be linked to the most recently ejected knot pair detected in each side of the jet. These observations show that the mass outflow rate remains a constant fraction of the mass accretion. A moderately lower mass outflow rate is found in the faster blue-shifted lobe, supporting the possibility that momentum ejection is conserved on each side of the jet. The evidence of frequent knot ejections indicates that this source is a good target for further monitoring to study the accretion-ejection connection.}


\maketitle
\section{Introduction}

Jets from young stellar systems are closely connected with the protoplanetary disk and with mass accretion onto the central star \citep{Calvet1997, JYSCabrit}. Magnetohydrodynamic (MHD) models of jet launching suggest that disk winds are launched from an extended region in the planet forming disk, and that the high-velocity (HV) jet represents the collimated inner part of the outflow which is launched from the inner 1-3 au of the disk \citep{Frank2014, Ray2021}. 

A signature of the connection between mass outflow and accretion is found in the knotty structures observed within jets, which are most likely due to internal shocks caused by changes in the outflow velocity. In recent years, several studies have demonstrated a time correlation between knot ejection and luminosity bursts due to episodic accretion, indicating that the knot structures trace the signature of variable accretion onto the star \citep{Takami2020}. Studying the variability in the jet therefore offers an avenue to probe the accretion-outflow connection. Multi-epoch observations show that the DG Tau and RW Aur jets exhibit significant changes in the frequency of knot ejections over a few decades which are linked to variable accretion, while changes in the velocity of ejected knots may relate to changes in the launching radius over time, with a faster outflow velocity reflecting a smaller launching radius closer to the inner disk edge \citep{Takami2023, Pyo2024}. However, the connection between knot launching and accretion bursts is not clearly understood: for example, whether all knots are linked to accretion bursts, or the precise relationship between the increase in mass outflow and accretion rate (e.g., whether smaller rises in accretion result in fainter observed knots).

Another aspect of this connection which is not yet understood is found in asymmetric jets. Many bipolar jets exhibit significant differences on either side of the source, in their morphology, brightness, velocity and gas properties such as electron density and temperature \citep{Hirth1997, Woitas2002}. In some targets only one jet lobe is detected, as in the HH 34 and Lk \Ha\ 321 jets \citep{Mundt1998, Reipurth2002, Coffey2004}. In some cases these asymmetries may be be extrinsic to the jet itself, i.e., caused by interaction with an inhomogenous circumstellar medium. For example, the RW Aur and DG Tau B jets show marked asymmetries in their appearance, but the mass outflow rate through each lobe is similar, suggesting that the underlying jet is symmetric \citep{Melnikov2009, Podio2011}. In the case of the asymmetric DG Tau jet, one side of the jet appears to interact with a clumpy, relatively dense medium, forming a bubble-like structure at 200 au from the star, while the other side propagates through a lower-density medium \citep{White2014b}.

In other cases, the asymmetry may be intrinsic to the launching mechanism, particularly where the asymmetric properties are traced close to the jet base; as for example in the DO Tau jet \citep{Erkal2021}. Measurements of the mass outflow rate in several asymmetric jets have produced mixed evidence as to whether the mass outflow rate through both lobes is equal, or in fact the faster jet lobe removes less mass. \citet{Ellerbroek2013} suggest that the mass outflow rate  may show a balance of linear momentum rather than mass loss. If the jet velocity is linked to the launching radius in the disk, understanding the balance of mass loss in asymmetric jets will enable us to understand both the rate of mass removal from each side of the disk, and the extent of the planet-forming region with which the jet interacts.

\begin{figure}
\centering
\includegraphics[width=9cm, trim= 0cm 0cm 0cm 0cm, clip=true]{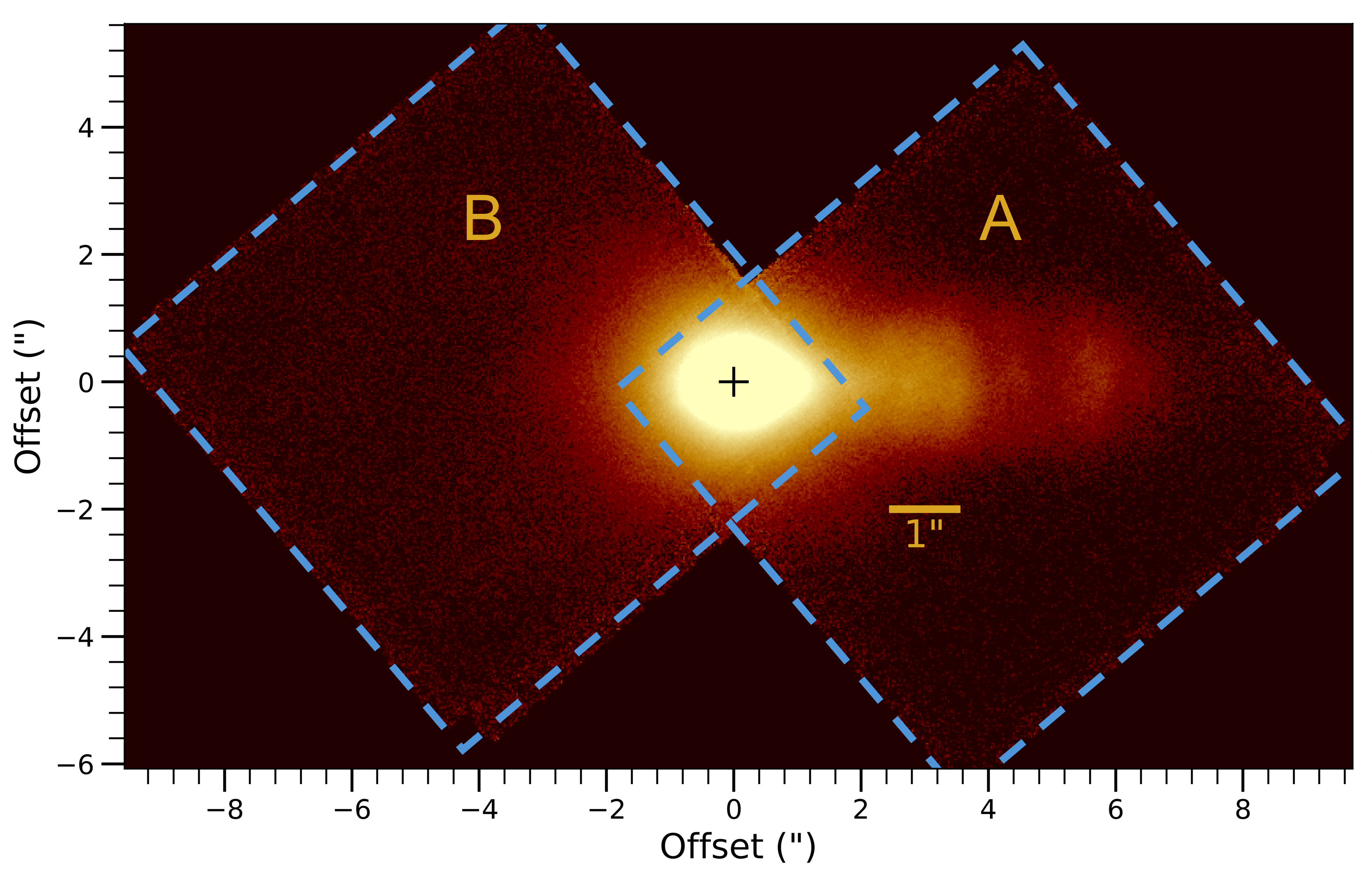}
\caption{The VLT/MUSE Narrow-Field Mode view of Th~28, showing the jet in H$\alpha$ emission before continuum subtraction. A black cross marks the source position (the peak of the continuum emission). The two fields of view A and B are highlighted, centred on the red-shifted (western) and blue-shifted (eastern) jet lobes, respectively.}

\label{fig:fov_diagram}     
\end{figure}

One such asymmetric jet source is the Classical T Tauri Star (CTTS) Th 28  (\object{ThA 15-28}; also referred to as Sz 102 or Krautter's Star), located in the Lupus III cloud (D $\sim$ 160 pc \citep{Dzib2018}). This source is very under-luminous (0.03 L$_{\odot}$, \citet{Mortier2011}) due to the edge-on orientation of the disk; however, significant mass accretion is indicated by the prominent outflow activity as well as the heavily veiled stellar spectrum. The key parameters of the star are summarised in Table 1 of \citet{Murphy2021}. Th 28 drives a bright bipolar jet (HH 228) which lies close to the plane of the sky. The jet axis is oriented in the east-west direction with PA $=$ +95$^{\circ}$ \citep{Graham1988}. 

The two jet lobes are strongly asymmetric in multiple respects. The blue-shifted lobe is significantly fainter in most FEL emission, and characterised by broad bow shocks as compared with the compact knot structures seen within 20\arcsec\ of the red-shifted lobe. The blue-shifted jet is also faster in both proper motions and radial velocity. Estimated proper motions in this lobe are 0\farcs47 \peryr compared with 0\farcs34-0\farcs4 in the red-shifted jet \citep{ Murphy2021}. The asymmetry is more pronounced in radial velocities, with \vrad\ = -80 \kms\ and +25 \kms\ in the blue- and red-shifted jets, respectively. Combining tangential and radial velocities yields estimated total jet velocities of \vjetb\ = 360 \kms\  and \vjetr\ =270 \kms\ \citep{Murphy2021}. The large asymmetry in radial velocity also implies a higher inclination of the blue-shifted lobe (twice that of the red-shifted lobe). Diagnostic analysis primarily using optical FEL emission lines consistently indicate higher temperatures and ionisation fractions in the blue-shifted lobe, as well as mostly yielding higher electron densities \eden\ \citep{ Coffey2010, Liu2014,Liu2021, Murphy2024}. The higher excitation in this jet is borne out by the extended [\oiii] emission seen in the blue-shifted lobe which is absent from the counter-jet. 

Importantly, this asymmetry appears to be intrinsic to the jet itself, as shown by \citet{Melnikov2023} who examine the [\feii] emission within the blue-shifted lobe and find no evidence of increased extinction as the jet becomes fainter; that is, the faintness of this jet is not due to propagating into a denser cloud on this side. The reason for the strong asymmetry in brightness is therefore unclear, but may be partially due to the wide spatial extent of the bow shocks in this lobe, resulting both in the deposition of the outflow energy over a wider spatial area, and in increased sideways loss of mass along the outflow. On the other hand, spectro-imaging of the two lobes show highly symmetric knot ejections on each side \citep{Murphy2021}, suggesting that the variability in the mass outflow is likewise symmetric. 

However, drawing a full comparison between the outflows at the base has proven difficult. While the morphological and kinematic properties of the red-shifted jet are well constrained due to its bright emission in a large number of optical lines, the fainter blue-shifted jet remains elusive. In part this is due to the lack of proper motion measurements for the knots in this lobe close to the star (within the field of view of previous spectro-imaging observations) and the more complex kinematics, with the jet emission lines showing a range of radial velocities which indicate strong bow shocks close to the source. Previous optical observations with the Wide Field Mode of the Multi Unit Spectroscopic Explorer on the Very Large Telescope (VLT/MUSE-WFM) \citep{Bacon2010} were able to map much of the extended emission in this lobe \citep{Murphy2021,Murphy2024}; however, this data had a seeing-limited spatial resolution (0\farcs9), resulting in suspected contamination of the inner blue-shifted jet by line emission from the brighter red-shifted lobe. As a result, these studies were unable to compare the jet properties and mass outflow rates in the inner 1-2\arcsec\ (160-320 au) where they are least affected by interaction with the surrounding environment. 

In this paper, we present new observations of the Th 28 jet obtained with the VLT/MUSE Narrow-Field Mode (NFM) at an effective angular resolution of 0\farcs12, enabled by the Adaptive Optics (AO) module. Our key aim is to compare the jet morphology, gas properties and mass outflow rates in the inner region of both jet lobes and to investigate the observed jet asymmetries. In Sect. \ref{section:target_obs} we present details of the target, observations and data reduction process. Sect. \ref{section:morphology} presents the spectro-images and morphological analysis of the data, while Sect. \ref{section:diagnostics} discusses the diagnostic analysis of the jet properties in each lobe, including the measurement of the mass accretion and outflow rates. A discussion of the results and implications is given in Sect. \ref{section:discussion} and our conclusions are summarised in Sect. \ref{section:conclusions}.

\section{Observations and data reduction}
\label{section:target_obs}

AO-assisted observations of the Th 28 jet were obtained in Narrow-Field Mode over two nights in June of 2023 and an additional night in January of 2024. The detector was rotated 45\ndeg\ in order to observe the maximum extent of the jet in each exposure. Exposures were divided into two fields of view (FoVs), A and B, centred on the western (red-shifted) and eastern (blue-shifted) jet lobes, with both fields overlapping around the central source as shown in Fig. \ref{fig:fov_diagram}.

\begin{table}
\centering
\caption[target]{\label{table:obs_details}Details of the VLT/MUSE NFM-AO observations.}
\vspace{0.1cm}
\renewcommand{\arraystretch}{1.4}
\centering
\begin{tabular}{{p{0.14\textwidth}<{\raggedright} p{0.07\textwidth}<{\raggedright} p{0.07\textwidth}<{\raggedright} p{0.1\textwidth}<{\raggedright} }}
\hline \hline
\multicolumn{4}{l}{\vspace{-0.4cm}} \\
Date &  FoV\footnote[1]{1} &  \makecell[l]{t$_{exp}$ \\(s)} & \makecell[l]{Sky resolution\footnote[2]{2} \\ (\arcsec) } \\
\hline
2023 June 15 & A & 860 & 0.096 \\
'' & A & 850 & 0.08 \\
'' & A & 850 & 0.096 \\
2023 June 17 & A & 1080 & 0.09 \\
'' & B & 750 & 0.089 \\
'' & B & 700 & 0.096 \\
2024 January 30 & B & 860 & 0.082 \\
'' & B & 850 & 0.082 \\
'' & B & 850 & 0.067 \\
\hline
\multicolumn{4}{l}{\vspace{-0.4cm}} \\
\multicolumn{4}{l}{\makecell[l]{Notes: (1) The field-of-view as denoted in Fig. \ref{fig:fov_diagram}. (2) The \\ effective angular resolution obtained after AO correction.}} \\

\hline
\end{tabular} 
\end{table}

Details of the individual exposures are shown in Table \ref{table:obs_details}. Exposures for FoV A were obtained during June with a total exposure time of 3640s. Two exposures for FoV B were obtained in the June period and the remainder were obtained during the January observation period with a total exposure time of 4010s. Average seeing during the exposures varied from 0\farcs7-0\farcs9 during the January observing block, compared with $\sim$ 0\farcs45 during the June block. The FoV B exposures therefore suffered from somewhat poorer observing conditions. However, the AO corrections result in an estimated effective angular resolution of 0\farcs07-0\farcs1 in all observing blocks. The spectral resolution of MUSE varies with wavelength from R = 1800 to 3600, with the corresponding velocity resolution ranging from 170~\kms\ (4750 \AA) to 80~\kms\ (9350 \AA). The key jet-tracing emission lines for this work fall in the range 6300-6750 \AA, with a velocity resolution of 125 \kms.

The data were reduced using the Reflex MUSE reduction pipeline (v. 2.8.9), including illumination correction, wavelength and flux calibration and exposure alignment. The pipeline applies a barycentric velocity correction, and we further correct all radial velocities quoted in this work for the LSR systemic velocity of +4 \kms\ \citep{Louvet2016}. 

The two FoVs combined form a mosaic as shown in Fig. \ref{fig:fov_diagram}. However, because the edge of each FoV cuts across the opposite lobe of the jet, this causes some contamination from edge pixels and potential distortion of the jet emission in these regions. To prevent this issue, each FoV was also reduced as a separate datacube covering one lobe of the jet. All results obtained from the combined mosaic were then compared with the same region in the corresponding eastern or western cube to ensure consistency and rule out the impact of the detector edge. In Sect. \ref{section:morphology} we examine the morphology and width of the inner jet regions using deconvolved spectro-images. To ensure the best spatial resolution in the blue-shifted jet, we deconvolve spectro-images of this FoV using only the June exposures (obtained during best conditions); however for the remainder of the analysis we combine all the exposures in this field for maximum S/N in the emission line flux.

In each datacube, the continuum emission was subtracted by fitting a polynomial function to the continuum level of the average spectrum at the source. This function was then scaled to match the continuum emission level at all other spaxels in the cube and subtracted from the data. The procedure also produced a datacube containing the estimated continuum emission at each spatial position. A small rotation is applied to the reduced datacubes to align the jet axis with the horizontal direction (using the \Ha\ emission in the brighter red-shifted jet as a reference, as in \citet{Murphy2021}).

The reduction pipeline calculates an estimate for the effective angular resolution after AO correction during each observation (quoted in Table \ref{table:obs_details}). These values are based on instrument calibration data rather than being measured directly from the observations. We therefore extracted continuum images at wavelengths of interest and fit the source continuum with the MAOPPY model of the MUSE long-exposure AO-corrected point spread function (PSF) \citep{Fetick2019}. This model includes a sharp Moffat core and wide wings, and takes into account both static aberrations and the cutoff frequency of the telescope. We find that the best-fit model yields an asymmetric fit: at 6500 \AA\ the approximate full width at half maximum (FWHM) of the core is 0\farcs25 in the horizontal direction (along the jet axis) and 0\farcs2 in the vertical direction. There is negligible variation between the mosaic image and the two individual FoV images, and between wavelength regions. The core width is also larger than the effective resolution of the observations; however we note that the source is surrounded by a compact, edge-on disk and subject to heavy obscuration. The measured width of the emission, therefore, should be considered as an upper limit to the true PSF width, and we adopt this as a conservative measure of the effective angular resolution before deconvolution.

\section{Morphology of the inner jet}
\label{section:morphology}

As estimated above, the NFM-AO assisted data has an effective on-sky angular resolution of 0\farcs2.  In comparison to the 2014 WFM observations (seeing-limited spatial resolution of ~0\farcs9), the NFM data therefore obtains a minimum improvement in spatial resolution of a factor $>$ 4.

To increase the resolution further and examine the morphology of the inner jet, the emission line spectro-images are deconvolved using a variant of the Richardson-Lucy (RL) algorithm implemented by the \texttt{sci-kit} Python package. To improve the image sharpening, we slightly modify the  \texttt{richardson$\_$lucy} function, to use the original image as the initial guess in the algorithm (rather than a flat array). We use as a PSF estimate the MAOPPY PSF model fitted to a continuum image extracted at the corresponding wavelength. The input PSF and jet images are normalised by the total flux and all negative values replaced with small near-zero (1 $\times$ 10$^{-37}$ \ergscm) flux to prevent artefacting. 

\begin{figure}[ht!]
\centering
\includegraphics[width=9cm, trim= 0cm 0cm 0cm 0cm, clip=true]{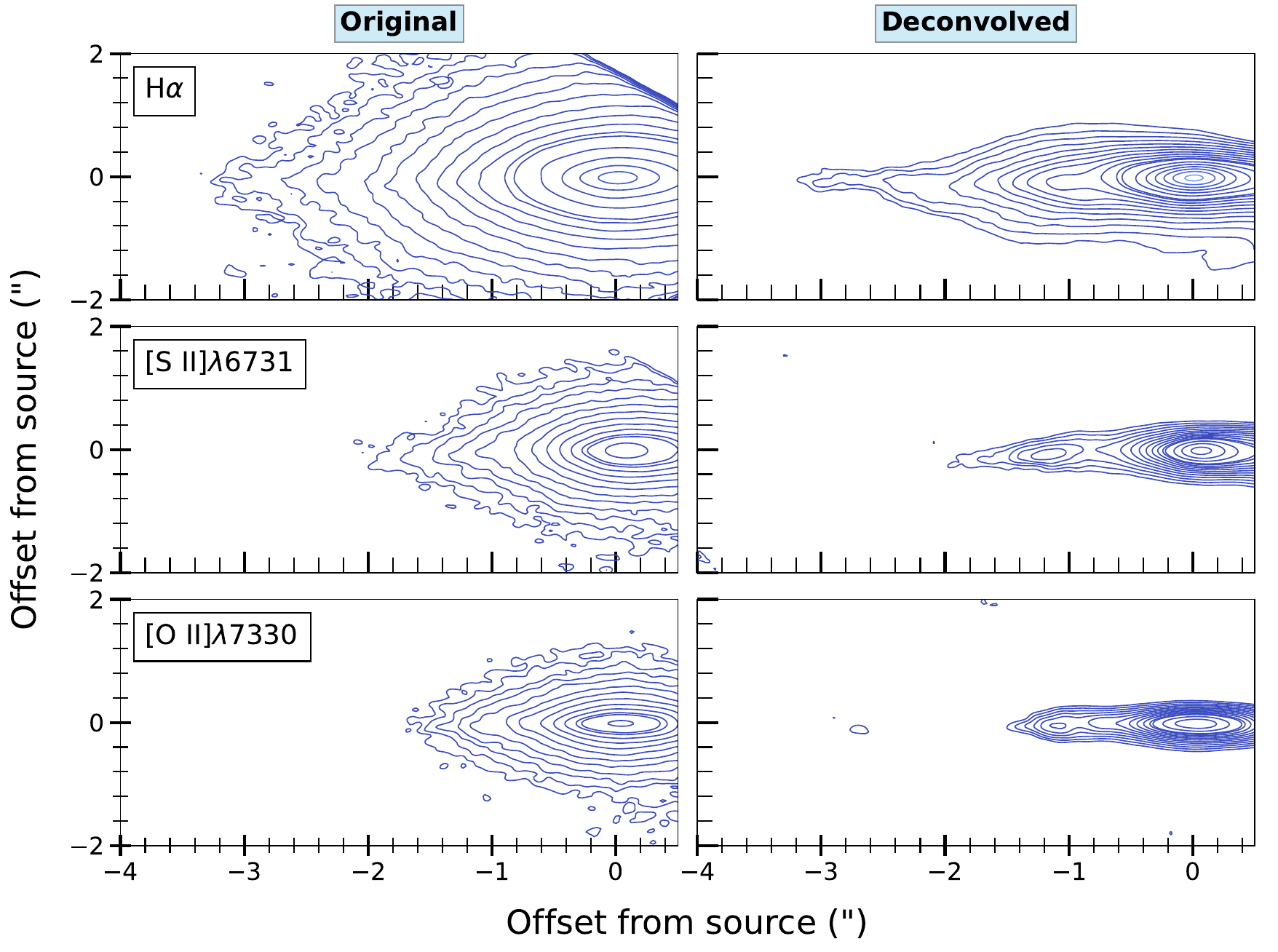}
\caption{\label{fig:decon_blue}Spectro-images of the blue-shifted jet channels (-140 \kms\ to +90 \kms) in Field B, before and after deconvolution. Contours are smoothed using a Gaussian filter (kernel $\sigma$ = 1.5) and start at 3-$\sigma$ of the background rms level pre-deconvolution, approximately 10 \bunit. The levels increase as a factor of $\sqrt{2.5}$. We note that the right-hand side of the image is distorted by the edge of the detector FoV. Additional emission lines are shown in Fig. \ref{fig:decon_ims_app}.}
\end{figure}

We similarly deconvolve the original continuum image with the PSF model, and find that at higher than 15 iterations the reduction of the core FWHM becomes negligible. At 15 iterations the FWHM is reduced to 0\farcs12, indicating an improvement in resolution by a factor of $\sim$2. We therefore use this as a benchmark to deconvolve the images.

\subsection{Knots in the inner jet}
\label{subsection:knot_ids}

\begin{figure}[t]
\centering
\includegraphics[width=9cm, trim= 0cm 0cm 0cm 0cm, clip=true]{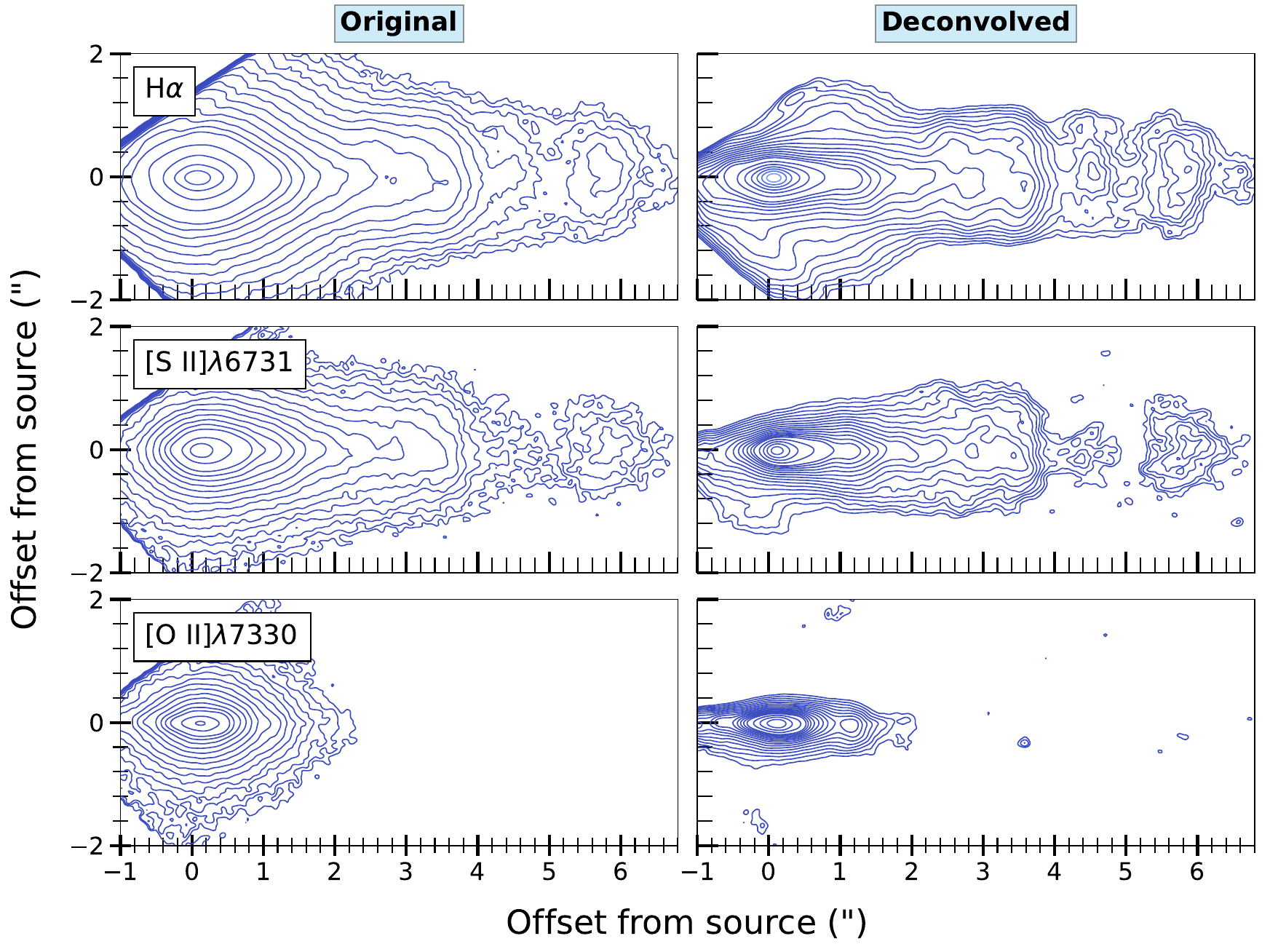}
\caption{\label{fig:decon_red} Spectro-images of the red-shifted jet channels (-90 \kms\ to +140 \kms) in Field A, before and after deconvolution. Contours are as in Fig. \ref{fig:decon_blue}, with starting levels of approximately 8 \bunit\ and increasing as a factor of $\sqrt{2}$. The left edge of the image is at the edge of the detector FoV. Additional emission lines are shown in Fig. \ref{fig:decon_ims_app}.}
\end{figure}

The blue- and red-shifted velocity channels (-140 \kms\ to +90 \kms\ and -90 \kms\ to +140 \kms, respectively) are shown for \Ha, \siib\ and [\oii]$\lambda$7330 in Figs. \ref{fig:decon_blue} and \ref{fig:decon_red}, which compare the images before and after deconvolution. Additional FEL spectro-images are shown in Fig. \ref{fig:decon_ims_app}. The blue-shifted jet is observed with an extent of up to -3\arcsec. The red-shifted jet is detected to an extent of approximately +6\arcsec, and shows clear knot structures located at approximately +3\arcsec, +6\arcsec\ and a faint knot at around +4.5\arcsec. The deconvolved images show additional peaks indicating a red-shifted knot at about +1\farcs2 (most visible in [\feii] and \Ha), a tentative peak at +2\arcsec, and suggest that the peak at 3\arcsec\ may be composed of two knot peaks at +2\farcs8 and +3\farcs2.

\begin{figure}[h!]
\centering
\includegraphics[width=9cm, trim= 0.3cm 0.2cm 0.2cm 0cm, clip=true]{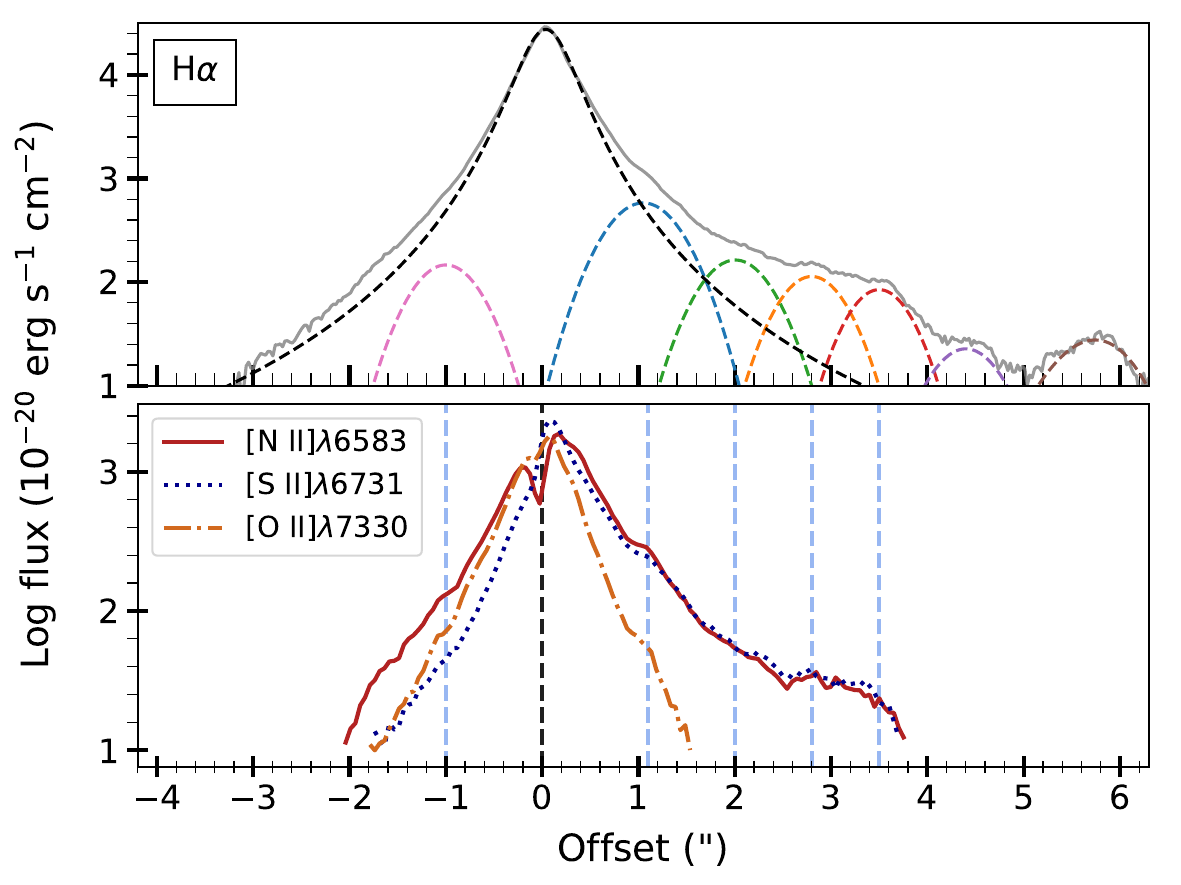}

\caption{\label{fig:flux_profiles}Selected flux profiles along the jet axis in the full mosaic (without deconvolution). Top panel: The \Ha\ profile (grey) without scattered emission subtraction, showing the components of the fitted model. The line emission at the source peak is fitted with a Moffat function (black dashed line) and knot peaks are fitted with Gaussian profiles (coloured dashed lines). Lower panels: Flux profiles in three key FELs after scattered emission subtraction (see Appendix \ref{section:scattered_sub}). The tentative and detected knot centres are marked with vertical blue lines, showing their correspondence to the features in the FEL profiles.}
\end{figure}

We check the unresolved knot positions in the inner jet by comparison with the flux profiles along the jet axis (before image deconvolution). To highlight the inner knot peaks in the FEL emission, we employ a procedure to remove some of the scattered line emission near the source, which is described in Appendix \ref{section:scattered_sub}. The flux profiles show peaks corresponding to the clearly resolved knot peaks as well as unresolved humps at approximately +1\arcsec\ in the red-shifted jet lobe and -1\arcsec\ in the blue-shifted jet. By fitting a combination of Gaussian peaks to the flux profile in \Ha, we estimate the peak locations which best reproduce the jet profile. Fig. \ref{fig:flux_profiles} shows the fitted peaks, and compares the estimated knot positions with the jet profiles in several FELs. Additional details of the fitted model are shown in Fig. \ref{fig:flux_profile_models}.

We find that the fitted peaks are generally consistent with the features observed in the FEL profiles after scattered emission subtraction, supporting the presence of these knot peaks. However, we note that the FEL profiles do not show distinct peaks at +2\farcs8 and +3\farcs2. Additionally, while the presence of a knot in the blue-shifted jet is supported by the fitted profile and by the small hump in the [\nii] and [\oii] profiles at $\sim$-1\arcsec, the structures in this lobe are not well resolved and the peak position is therefore very uncertain. The proper motions and dynamical timescales of the jet knots are discussed further in Sect. \ref{section:discussion}.

The knot line profiles were sampled in \Ha\ as well as \OIa, \NII, \SII, \Feiia\ and \Feiib\ for comparison. Table \ref{table:knot_ids} lists the resulting knot locations as well as the \siib\ radial velocities (\vrad) and proper motions. The full \vrad\ and FWHM measurements are given in Tables \ref{table:knot_vrad} and \ref{table:knot_fwhms}. Given the velocity resolution of 125 \kms, the line widths for knots within 1-2\arcsec\ of the source are well resolved, while for the knots in the extended red-shifted jet at separations $>$2\arcsec\ these are marginally resolved at most. The line profiles in the blue-shifted jet will be discussed in Sect. \ref{subsection:bluejet_knots}. In the red-shifted lobe, \vrad\ is very consistent for all the knot positions in \Ha\ and FEL measurements, with the exception of \OIa\ within 2\arcsec\ which peaks approximately 3-4 \kms\ below other emission lines. This may be due to a small contribution from the Low-Velocity Component close to the source. In general, \vrad\ increases slowly along the jet from +18 to +25 \kms; this effect may be partially due to the knots fading with distance from the source, so that only the bright high-velocity peak is still detected in more distant knots. The FWHM of the line profiles is also very consistent between FELs, ranging from 200 \kms\ at the first knot to 130 \kms\ at the most distant knot position. The \Ha\ profiles have a slightly larger FWHM (approximately 20 \kms\ higher than the FEL values).

\begin{table}
\centering
\renewcommand{\arraystretch}{1.5}
\caption[target]{\label{table:knot_ids}Knots detected in the NFM observations.}
\vspace{0.1cm}
   \begin{tabular}{{p{0.08\textwidth}<{\raggedright} p{0.08\textwidth}<{\raggedright} p{0.1\textwidth}<{\raggedright} p{0.07\textwidth}<{\raggedright}}}
    \hline \hline
    \makecell[l]{Offset \\ (\arcsec)} & \makecell[l]{v$_{rad}$\footnote[1]{1} \\ (\kms)}  & \makecell[l]{ PM \\(\arcsec\ \peryr)} & \makecell[l]{ t$_{dyn}$ \\(yr)}\\
    \hline
    - 1.2 \tiny{$\pm$ 0.2} & -40-70 &  0.47\footnote[2]{2} \tiny{$\pm$ 0.1} & 2.6 \tiny{$\pm$ 1} \\
    \hline
    + 1.2 \tiny{$\pm$ 0.1} & 20 &  0.34\footnote[2]{2} \tiny{$\pm$ 0.05} & 3.5 \tiny{$\pm$ 1}\\
    + 2.0 \tiny{$\pm$ 0.1} & 20 &  0.34\footnote[2]{2} \tiny{$\pm$ 0.05} & 5.9 \tiny{$\pm$ 1} \\
    + 2.8 \tiny{$\pm$ 0.1} & 22 & 0.34\footnote[2]{2} \tiny{$\pm$ 0.05} & 8.2 \tiny{$\pm$ 2} \\
    + 3.5 \tiny{$\pm$ 0.1} & 21 &  0.34\footnote[2]{2} \tiny{$\pm$ 0.05} & 10.3 \tiny{$\pm$ 2} \\
    + 4.5 \tiny{$\pm$ 0.1} & 24 & 0.41\footnote[3]{3}  \tiny{$\pm$ 0.1} & 11.0 \tiny{$\pm$ 3} \\
    + 5.9 \tiny{$\pm$ 0.1} & 25 & 0.38\footnote[3]{3}\tiny{$\pm$ 0.05} & 15.5 \tiny{$\pm$ 2} \\
    \hline 
    \multicolumn{4}{l}{\vspace{-0.4cm} } \\
    \multicolumn{4}{l}{ \makecell[l]{Notes: (1) Measured from the \siib\ line profile. \\ (2) Proper motions  from \citet{Murphy2021}. \\ (3) From this work.} } \\
    \multicolumn{4}{l}{\vspace{-0.5cm} } \\
    \hline
    \end{tabular}
\end{table}

\subsection{Knot properties in the blue-shifted jet}
\label{subsection:bluejet_knots}

A key limitation in the previous optical diagnostic study \citep{Murphy2024} was the difficulty in separating emission from the red- and blue-shifted jets, particularly given the seeing-limited angular resolution of the datasets and the faintness of the blue-shifted side. In the NFM observations, the improved resolution should minimise the contamination between lobes. However, we also consider the possibility of scattered emission in the inner 1\arcsec\ of the jets due to the compact edge-on disk. Examination of the continuum profiles (Fig. \ref{fig:continuum_model_cross_section}) shows that the emission drops substantially by around +/- 0\farcs5 from the source, making it likely that scattered jet emission is also sharply reduced by this point.  

We further examine the line profiles on each side of the jet to check where the emission can be distinguished in the two lobes. Fig. \ref{fig:profiles_0_2as} shows the key FELs and \Ha\ sampled at +/- 0\farcs2, 0\farcs4 and 0\farcs6 from the source. We find that high excitation lines of [\oiii] and [\nii] show a clear separation in the profile peaks at offsets of +/- 0\farcs2 from the source, whereas the [\sii] and \Ha\ profiles are less clear, but show clearly distinct profiles from 0\farcs4-0\farcs6 from the source. The \OIa\ line shows very similar line profiles at all positions since it is centred at low velocities on both sides; however, the profiles on each side become more distinct at least 0\farcs6 from the source. On this basis, we estimate that we can sample the jet lobes separately from distances of at least +/- 0\farcs5 from the source position. This is a substantial improvement over the previous study where the minimum separation from the jet base was +/- 1\arcsec\ and significant contamination from the red-shifted jet could not be ruled out.

The line profiles in the blue-shifted jet are difficult to interpret due to the low MUSE velocity resolution, but show multiple components in \Ha\ and [\sii] including enhanced high-velocity emission. The [\nii] and [\oiii] lines are well fitted by a single Gaussian component, and yield consistent \vrad\ of -40 to -60 \kms\ and -80 \kms\ along the jet, respectively.  Fitting a single Gaussian to the \Ha\ profile yields a central \vrad\ -14 to -18 \kms\ (though we note that the profile shape is not very well fitted by a single peak).  The [\sii] lines show low velocities of $\leq$ -10 \kms\ within 1\arcsec\ of the source, with an increase to -25 to -30 \kms\ at the knot position. Within this inner 1\arcsec\ the [\sii] peak velocity may be reduced by contributions from the low-velocity shock wings. The [\oi] line shows a narrow red-shifted peak at $<$ 5 \kms\ along the full extent of this lobe; this is close to rest velocity given the MUSE resolution.

Taking \vrad\ from the [\oiii] and [\nii] peaks, with estimated proper motions of 0\farcs4 \peryr\ in this jet (\vtan\ = 340 kms) yields an inclination of -12 to -14\ndeg, similar to that found previously and more than twice the inclination of the red-shifted jet. However, using \vrad\ of -18 to -30 \kms\ from the \Ha\ and [\sii] peaks yields inclination $\sim$ 4\ndeg, similar to the inner red-shifted jet. The velocity FWHMs in this jet are quite broad, ranging from 250-270 \kms\ in the [\sii] lines to 300 kms in [\oiii] and 360 \kms\ in \Ha\ (see Fig. \ref{fig:knot_profiles} and Table \ref{table:knot_fwhms}). We note that [\oi] has a narrower FWHM of 210 \kms, similar to that in the red-shifted jet.

\subsection{Jet width}
\label{subsection:jet_width}

\begin{figure}[h!]
\centering
\includegraphics[width=9cm, trim= 0cm 0cm 0cm 0cm, clip=true]{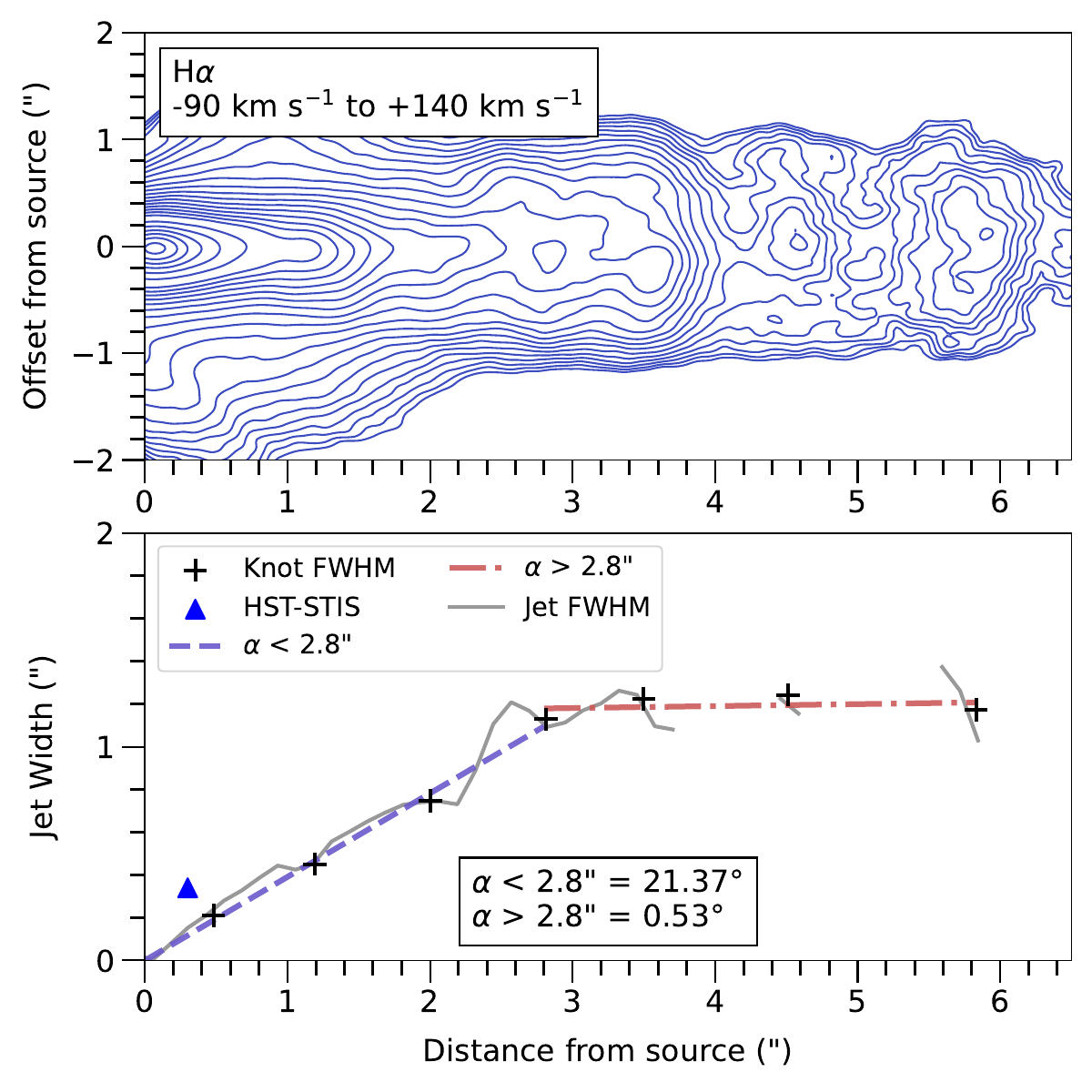}
\includegraphics[width=9cm, trim= 0cm 0cm 0cm 0cm, clip=true]{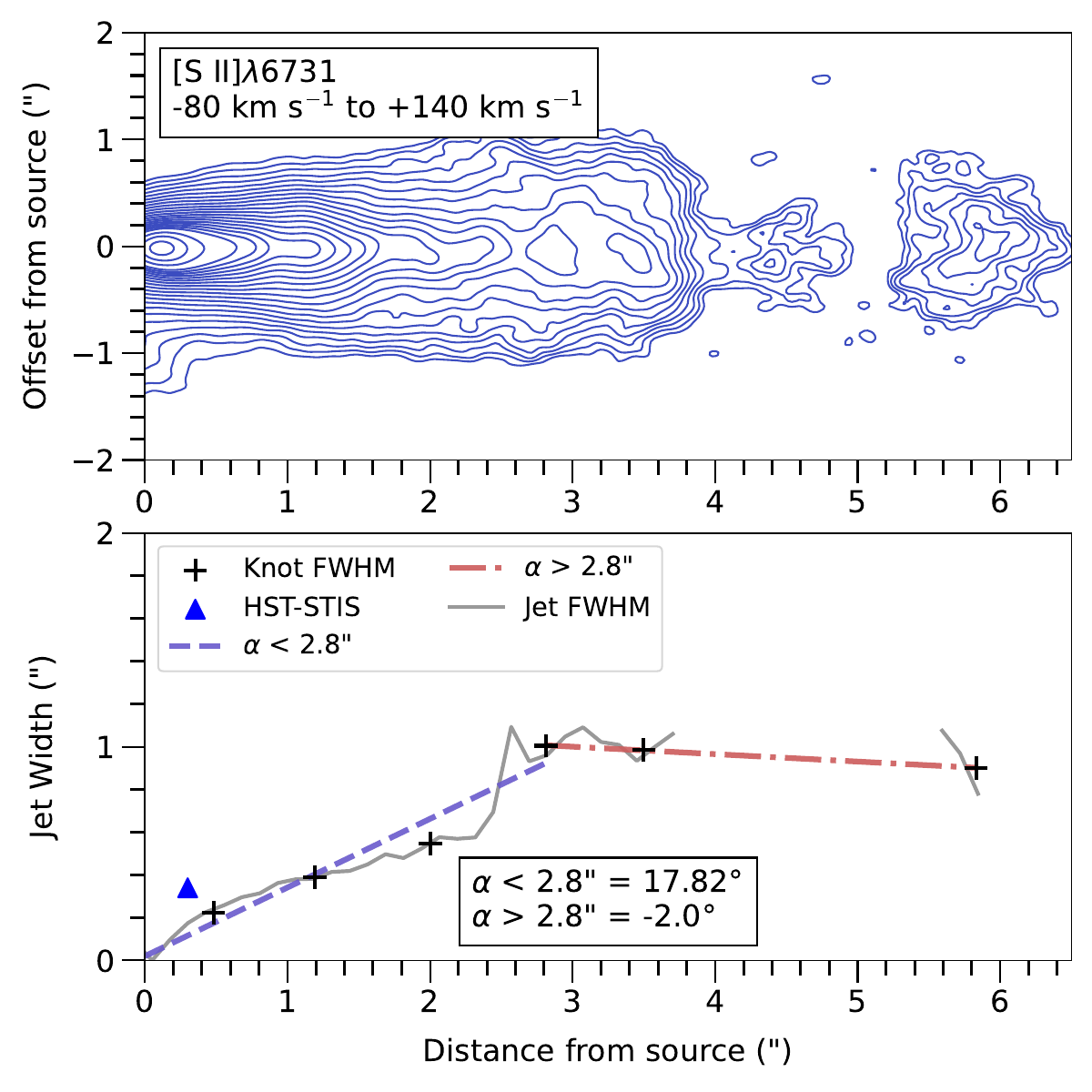}
\caption{\label{fig:jet_width_fitting}Deconvolved jet widths and opening angles $\alpha$ fitted across the full red-shifted jet velocity range; widths are fitted at +0\farcs5 from the source and at the knot positions. Over-plotted is the average jet width obtained for this lobe from HST-STIS optical data \citep{Coffey2004,Coffey2007}.}
\end{figure}

The jet width was investigated using deconvolved spectro-images in \Ha, [\nii], \SII, [\oii]$\lambda$7330 and \Feiib\ for each of the red- and blue-shifted jets. The spectro-images are integrated over a range of -90 \kms\ to +140 \kms\ in the red-shifted jet and -140 \kms\ to +90 \kms\ in the blue-shifted jet (the same as shown in Figs. \ref{fig:decon_blue} and \ref{fig:decon_red}). The spatial cross-section of the jet was sampled at each of the identified knot positions; the spatial full width at half maximum (FWHM) was then estimated by fitting a Moffat profile (in the red-shifted jet a Gaussian profile was used at offsets $>$2\arcsec\ as this provided a better fit). The corresponding PSF width was estimated from a deconvolved continuum image (FWHM = 0\farcs122); this PSF width was then subtracted in quadrature from the knot FWHMs to recover the intrinsic jet width. Examples are shown in Fig. \ref{fig:jet_width_fitting} for the red-shifted jet and in Fig. \ref{fig:bluejet_width_fitting} for the blue-shifted jet.

In the red-shifted jet, the width is resolved to +0\farcs2 from the source. The jet widens from 30-70 au at +0\farcs5 to the first knot position at +1\farcs2, and thereafter widens slowly. At the position of the large knot at +2\farcs8, the width increases sharply to $\sim$160 au in FEL emission and $\sim$190 au in \Ha. These widths remain stable in the knots at +3\farcs5 and +5\farcs8. Within 3\arcsec, we obtain an opening angle of 20\ndeg\ in both the FELs and \Ha.  Two exceptions are the [\oii] emission, which traces higher-excitation regions and which has a slightly narrower opening angle of 15\ndeg, and [\feii], which traces the highest density regions and which shows a much narrower opening angle of 7\ndeg. Both of these lines are observed only within 1\farcs5 of the source and so cannot be used to measure the jet width at larger offsets.

Beyond 3\arcsec\ from the source, the opening angle becomes very small (0.5\ndeg\ in \Ha) or even negative due to a decrease in width at the most distant knot at +6\arcsec. Given the arc shaped structure of the knot, this may be a natural consequence of the broader wings of the emission fading in the older knot.

Measuring the width of the blue-shifted jet is more difficult due to the shorter extent of the detected emission, and we do not find extended [\feii] in the deconvolved spectro-images. In the other lines, we measure the jet widths at offsets of -0\farcs5, -1\arcsec\ (the approximate position of the closest knot) and in the brighter \Ha\ emission we also measure the width at -1\farcs5. Despite the diffuse emission in this jet, the deconvolved widths are comparable to those of the red-shifted jet at similar distances, increasing from approximately 30-80 au near the jet base. In both the [\nii] and [\oii] images, the emission is narrower in width than the [\sii] and \Ha\ emission, but has a sharp opening angle of 12-13\ndeg\ compared with 6-8 in [\sii]. The \Ha\ images show an initial opening angle of 25\ndeg\ before the first knot position at -1\arcsec, with the jet width appearing to stabilise beyond this point ($\alpha$ = 1.2\ndeg) similar to the red-shifted jet. 

These wide opening angles in the inner jet are consistent with those found by \citet{Melnikov2023}, who estimate $\alpha$ = 28\ndeg\ in the inner 1\arcsec\ of both lobes based on [\feii] spectro-images. They similarly show the jet width stabilising after the position of the first detected knot, which suggests that on larger scales the jet width is dominated by the shock structures.  As in their measurements, we find a jump in the  red-shifted jet width behind the knot position, followed by a decrease at the knot itself; this is naturally due to the narrower width of the bright emission peak and does not reflect a change in the true jet width.

\section{Emission line diagnostics at the jet base}
\label{section:diagnostics}

We examine the emission line ratios measured within the inner bright portion of the jet (up to -2\arcsec\ in the blue-shifted lobe and +3\farcs5 in the red-shifted lobe). All fluxes are sampled along the jet axis from a 0\farcs4 $\times$ 0\farcs4 box, with a radius matching the unconvolved PSF width. 

We further estimate the physical conditions within each of the jet lobes using key diagnostic emission line ratios as detailed in \citet{BE99} (hereafter ‘the BE method’). The flux ratios between these lines provide direct tracers of the gas properties which can be cross-referenced between emission lines which are sensitive to different regimes of electron density \eden. A key goal is to estimate the mass outflow rate through each lobe, which requires estimates of the electron density, ionisation fraction and hence the hydrogen density (\nh\ = $n_{\mathrm{e}}/x_{\mathrm{e}}$). The BE method combines emission line ratios of O, S and N to obtain estimates of these parameters. 

In Sect. \ref{subsection:vshock} we examine line ratios tracing the shock velocities within the inner jet region; in Sect. \ref{subsection:be_ratios} we present the results obtained with the BE method near the jet base; and in Sect. \ref{subsection:oii_ratios} we extend these diagnostics by also considering the line ratios using [\oii]$\lambda$7320 and $\lambda$7330. Finally, in Sects. \ref{subsection:macc} and \ref{subsection:mout} we estimate the mass accretion rate at the source and the mass outflow rate through the jet, respectively.

\subsection{Shock velocities in the inner jets}
\label{subsection:vshock}

\begin{figure}
\centering
\includegraphics[width=8.5cm, trim= 0.2cm 0.cm 0cm 0cm, clip=true]{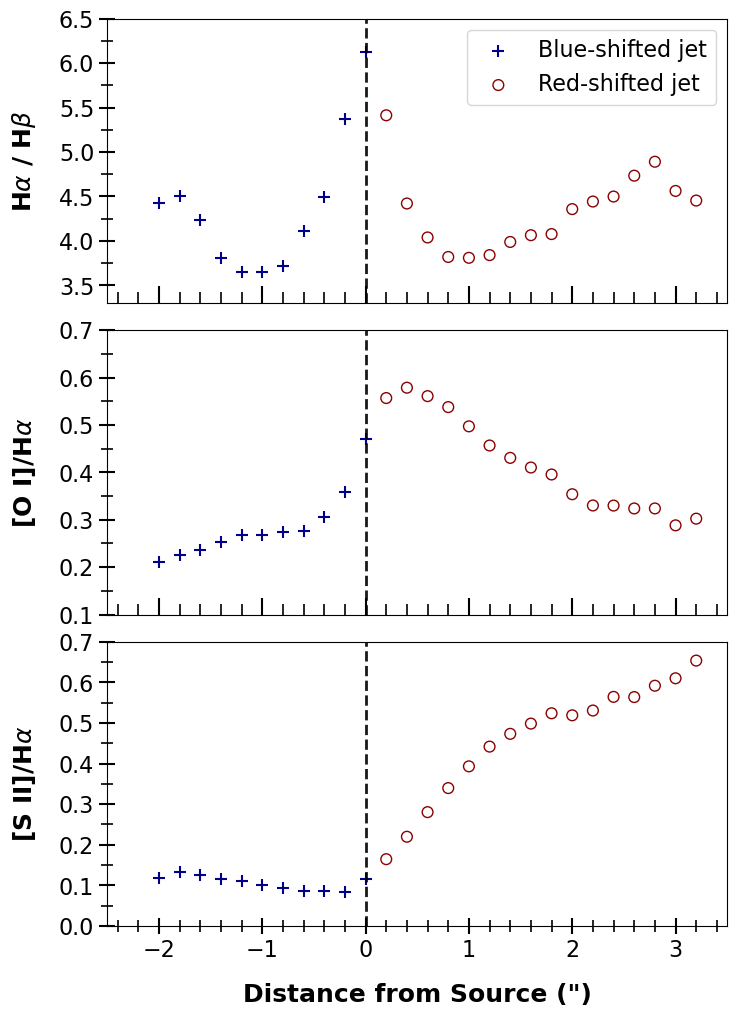}
\caption{\label{fig:flux_ratios_shocks}Emission line ratios which trace shock velocities in the inner jet regions. No extinction correction is applied as the \OI, \SII\ and \Ha\ lines are close in wavelength, and we use the \Ha/\Hb\ ratio to estimate the extinction (Sect. \ref{subsection:extinction}). The source position is marked by the green dashed line.}
\end{figure}

Fig. \ref{fig:flux_ratios_shocks} shows three line ratios which trace shock velocities (\vshock) within the jet. The [\oi]/\Ha\ and [\sii]/\Ha\ ratios use lines with similar wavelengths and are minimally affected by extinction, while the \Ha/\Hb\ ratio is sensitive to extinction as well as \vshock. We compare the measured ratios with the theoretical curves calculated by \citet{Hartigan1994}. In the blue-shifted lobe we assume \eden\ $>$ 10$^{4}$ \percm\ based on the high electron densities obtained in \citet{Murphy2024}. The [\oi]/\Ha\  and \sii/\Ha\ ratios in this region yield values of 0.2-0.25 and $\sim$0.1, respectively. These values are both consistent with shock velocities $>80-100$ \kms, as expected given the prominent [\oiii] emission in this jet.

In the red-shifted lobe, we take \eden\ $>$ 10$^{4}$ \percm\ within 1\arcsec\ of the source where the gas density is highest, and \eden\ $\sim$ 10$^{3}$ \percm\ at larger offsets. The [\oi]/\Ha\ ratio decreases along the jet axis, from $>$0.55 close to the source, to 0.3 at the knot located at 3\arcsec, values which are consistent with shock velocities of 40-60 \kms. The [\sii]/\Ha\ ratio yields values of 0.2-0.35 within 1\arcsec\ of the source, suggesting \vshock\ of 50-70 \kms, increasing to 0.6 at +3\arcsec\ (\vshock\ $\sim$ 45 \kms). The overall values of these ratios are small, suggesting fast shocks which are consistent with the [\oiii] emission near the base of this lobe.

\subsection{Extinction}
\label{subsection:extinction}

The estimated extinction towards Th 28 varies between 1-6 mag in previous studies. \citet{Melnikov2023} estimate \av\ = 0-6 mag in the red-shifted jet based on the [\feii] line ratios and 6 mag in both lobes and at the source position based on the \mh emission, while \citet{France2012} find \av\ = 1.1 mag based on pre-main-sequence evolutionary tracks. Based on the \Ha/\Hb\ decrement obtained in the MUSE-WFM data, \citet{Murphy2024} estimated \av\ = 2.5 mag at the source position (+/- 1\arcsec), whereas the [\feii]1.64 $\mu$m/[\feii]1.32 $\mu$m and [\feii]1.64 $\mu$m/[\feii]1.26 $\mu$m ratios obtained from X-Shooter spectra are consistent with an average Av = 1.26 mag in this region. However, the value obtained from the \Ha/\Hb\ decrement assumes an intrinsic ratio value corresponding to the Case B recombination limit (i.e., both lines are in optically thick emission at the source), and this assumption may be inapplicable to T Tauri sources due to the contribution of multiple emission components from different regions \citep{Antoniucci2017}. The same study also sampled an array of emission lines tracing mass accretion, and found that de-reddening the line fluxes for \av\ = 2.5 mag resulted in significant over-correction (i.e., a systematic decrease in luminosity with increasing wavelength). This further supports the lower estimate for extinction at the source. The \Ha/\Hb\ ratio is also sensitive to shock velocities in the jet \citep{Hartigan1994}, and \citet{Murphy2024} find \av\ = 0-0.5 mag for \vshock\ = 40 \kms\ in the red-shifted jet, and 1-1.5 mag in the blue-shifted jet within 3\arcsec.

In the NFM data, we again examine the \Ha/\Hb\ ratio (Fig. \ref{fig:flux_ratios_shocks}). For Lupus, we adopt the extinction curve RV = 5.5 \citep{Evans2009, Mortier2011} and apply the extinction law of \citet{Cardelli1989}. As in the WFM data, the ratio varies significantly along the jet, with the highest peak at the source position. If we again assume Case B conditions in this region, we find a peak value of \av\ = 3 mag.  In the blue-shifted lobe, we find values of $\sim$3.5 at offsets between -0\farcs5 and 1\farcs5, corresponding to \av\ = 1 mag for \vshock\ $>$ 90 \kms. In the red-shifted lobe, values increase from 3.6 at the base to 5.0 at 3\arcsec. If we assume \vshock\ = 70 \kms\ in the region up to the first knot at +1\farcs2, we obtain \av\ = 1 mag in this region as well. In the remaining extent of the red-shifted jet, we find \av\ = 0-0.6 mag for \vshock\ of approximately 40 \kms. If the shock velocities in this lobe are below the values we estimate, the derived extinction values will accordingly be reduced (as the intrinsic value of \Ha/\Hb\ will be increased). Since \citet{Melnikov2023} calculated their extinction values assuming RV = 3.1, we also considered this case; taking the same RV value our extinction estimates would be even lower, with a maximum \av\ at the source of 2 mag and 0.6 mag in the inner red- and blue-shifted lobes.

Given these uncertainties and the low estimated extinction in both jet lobes, and as most of the line ratios examined in this work are similar in wavelength, we do not apply an extinction correction in most sections of this analysis. We make note of two exceptions, however. In Sect. \ref{subsection:oii_ratios}, we examine line ratios of [\sii], [\oi] and [\oii] using emission lines which differ by $\sim$1000 \AA; we therefore consider these ratios for \av\ between 0-2 mag as a representative estimate for both jet lobes. In Sect. \ref{subsection:macc}, we estimate the mass accretion rate using the emission lines sampled at the source position, where extinction is likely highest. To facilitate comparison with the previous analysis in \citet{Murphy2024} and avoid the over-correction described above, we apply the lower correction for \av\ = 1.26 mag to the sampled line fluxes.

\subsection{Diagnostic line ratios}
\label{subsection:be_ratios}

\begin{figure*}
	\centering
	\includegraphics[width=15cm, trim= 0cm 0.cm 0cm 0cm, clip=true]{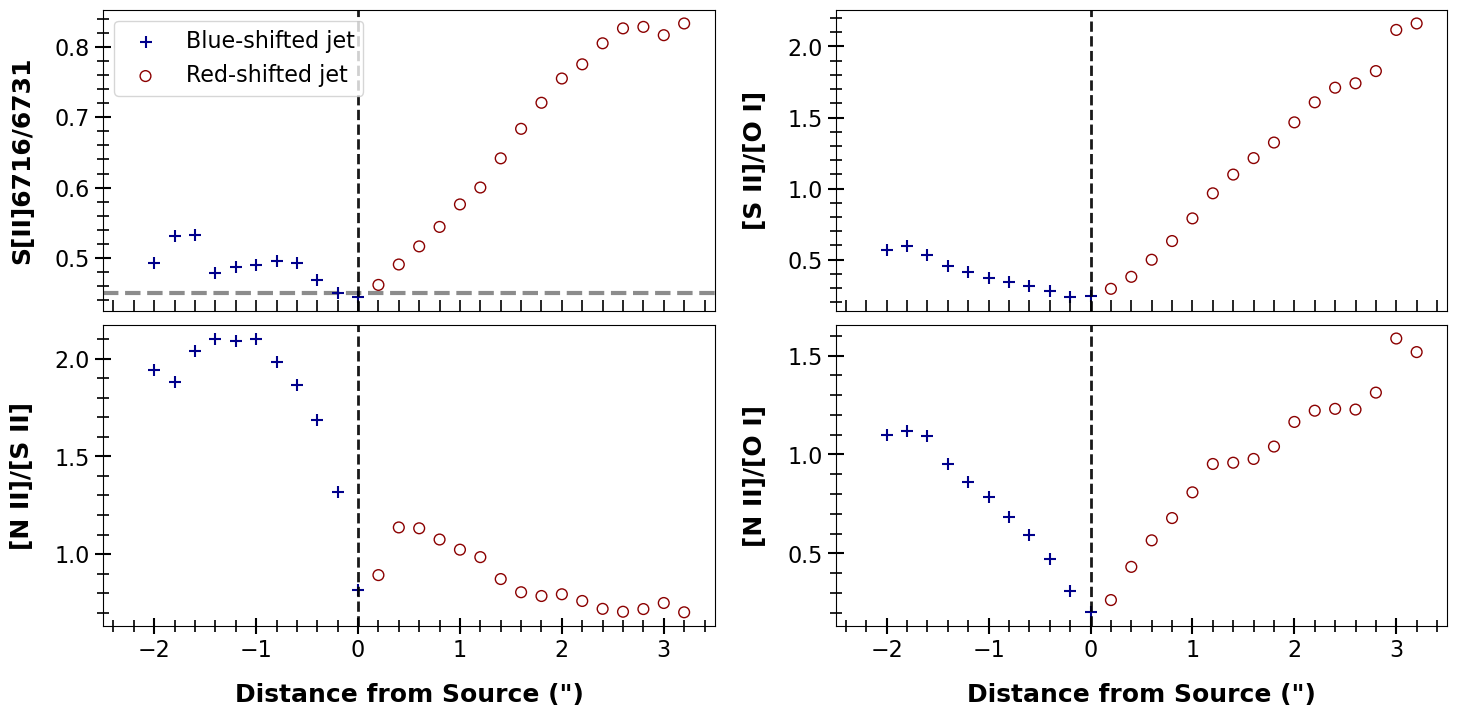}
    \includegraphics[width=15cm, trim= 0cm 0cm 0cm 0cm, clip=true]{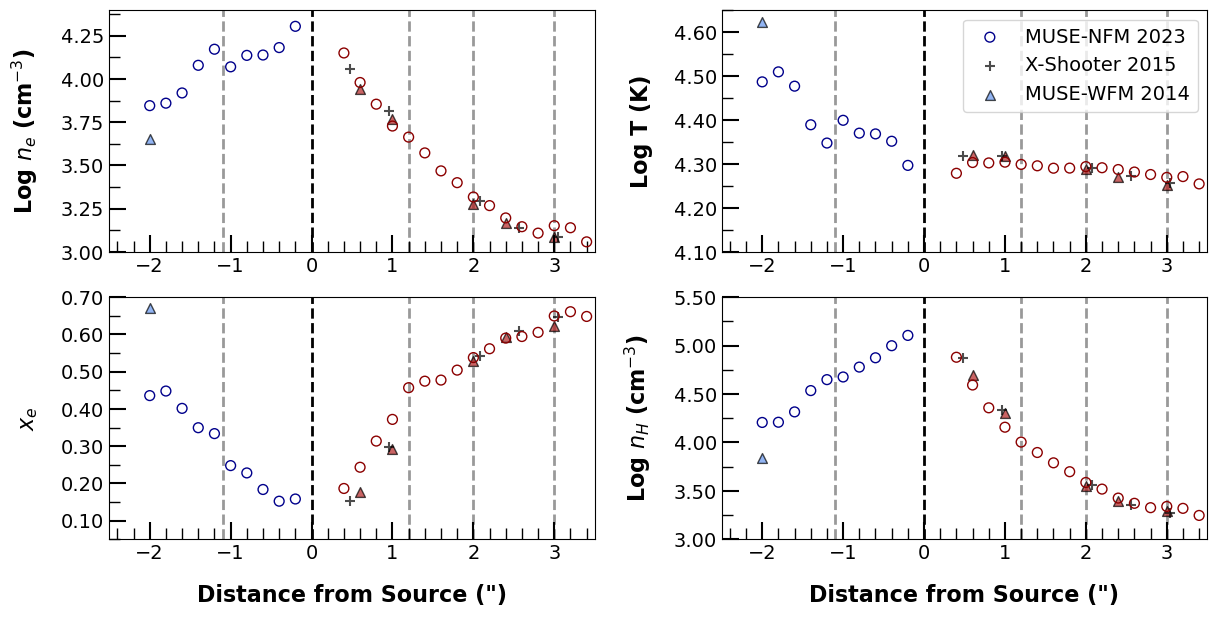}
	\caption{\label{fig:flux_ratios_BE}Top: Flux ratios of FELs used in the BE diagnostics, measured along the jet axis. The upper left panel shows \siirat; upper right panel shows [\sii]$\lambda\lambda$(6716+6731)/[\oi]$\lambda\lambda$(6300+6363); lower left panel shows [\nii]$\lambda\lambda$(6548+6583)/[\sii]$\lambda\lambda$(6716+6731); and lower right panel shows [\nii]$\lambda\lambda$(6548+6583)/[\oi]$\lambda\lambda$(6300+6363). Fluxes are sampled from the inner +/- 0\farcs2 around the axis centre to highlight the properties of the central flow. The blue- and red-shifted outflow directions are left and right of the source position (black dashed line) respectively. Bottom: the derived gas properties obtained using the BE method to combine these lines. Here the line fluxes are sampled across the full jet width (before deconvolution). Overplotted are the corresponding measurements from the previous MUSE-WFM and X-Shooter observations \citep{Murphy2024}. Grey dashed lines mark the knot positions.}
\end{figure*}

In Fig. \ref{fig:flux_ratios_BE}, we show four key line ratios tracing \eden, \Te\ and \xe\ within the jet. The ratio \siirat\ is a well-known tracer of \eden\ below the critical density of $\sim$2 $\times$ 10$^{4}$ \percm. At electron temperatures $>$ 5000 K, the ratios [\sii]$\lambda\lambda$(6716+6731)/[\oi]$\lambda\lambda$(6300+6363) (hereafter the [\sii]/[\oi] ratio), [\nii]$\lambda\lambda$(6548+6583)/[\sii]$\lambda\lambda$(6716+6731) (the [\nii]/[\sii] ratio), and [\nii]$\lambda\lambda$(6548+6583)/[\oi]$\lambda\lambda$(6300+6363) (the [\nii]/[\oi] ratio) are each tracers of \Te\ and \xe\ in the jet, i.e. excitation \citep{Hartigan1994, Podio2011}. In regions above the [\sii] high-density limit (HDL), the [\sii]/[\oi] ratio also has a strong dependence on \eden. The [\nii]/[\sii] ratio primarily depends on \xe\ and \Te, but shows a weak positive dependence on \eden\ between densities 10$^{4}$ to 10$^{6}$ \percm, and is a good tracer of high-excitation shock regions \citep{Hartigan2007}. The [\nii]/[\oi] ratio is most sensitive to  \xe\ and \Te\ below 2 $\times~10^{4}$ K in low-density regions (\eden\ $<$ 10$^{4}$ \percm) and weakly dependent on \eden\  at higher densities. These four line ratios form the basis of the BE diagnostic method which we use to examine the jet gas properties, and we therefore wish to examine them directly.

From Fig. \ref{fig:flux_ratios_BE}, we find that the [\sii]/[\oi] ratio increases smoothly with distance from the source on both sides of the jet, with very slight increases if any at the positions of the red-shifted knots.  The [\nii]/[\sii] ratio shows a broad hump centred at about -1\arcsec\ in the blue-shifted jet and a steady decline along with the red-shifted jet with small bumps at the knot positions. The [\nii]/[\oi] ratio also increases along the axis on both sides, with a generally smooth increase in the blue-shifted lobe and much more distinct increases at the knot positions in the red-shifted lobe.  Two points can be highlighted from these results. First, the knot positions identified by the spectro-images and flux profiles are reflected by jumps in excitation at these positions. Second, the blue-shifted jet appears to be above the [\sii] HDL and these line ratios will therefore have a stronger dependence on \eden\ in this lobe, reflecting higher overall densities. Although the [\nii]/[\sii] ratio shows a peak at about the position of the possible blue-shifted knot, this is broader and much less distinct than the red-shifted knots. This peak is much higher than the values obtained in the other jet, however, highlighting the high excitation in this lobe.

We estimate the physical conditions within the jets using the BE method, which utilises the jet tracing FELs [\oi]$\lambda$$\lambda$6300, 6363, [\nii]$\lambda$$\lambda$6548, 6583, and [\sii]$\lambda$$\lambda$6717, 6731. Here we make use of an updated numerical code which applies updated abundances and collision strengths and accounts for the effect of \Te\ dependencies when calculating \eden\ values, making it more suitable for regions of high ionisation and excitation. The analysis assumes solar photospheric abundances of S, N and O taken from \citet{Asplund2005}. Both the code and the related uncertainties of the method are discussed in \citet{Murphy2024} (see also further discussion in \citet{Podio2006b,Melnikov2008}). 

A caveat which must be highlighted is that the BE method should be applied in regions where [\oi] emission dominates over higher ionisation lines, i.e. where [\oii] and [\oiii] emission is negligible. As discussed in \citet{Murphy2024}, the blue-shifted lobe exhibits significant [\oiii] emission; in Fig. \ref{fig:o_profiles} we compare the flux along the jet axis from the [\oi]$\lambda$6300, [\oii] ]$\lambda$7330 and [\oiii] ]$\lambda$5007 lines. In the blue-shifted lobe, the [\oiii] emission dominates over [\oi] from approximately -0\farcs5. We report the results from this lobe for comparison, but emphasise that the ionisation fraction in this region will not be well constrained as the ionisation balance is not well known.

The estimated gas properties are shown in Fig. \ref{fig:flux_ratios_BE} with the corresponding estimates from the 2014-2015 MUSE-WFM and X-Shooter data overplotted \citep{Murphy2024}. The NFM results in the red-shifted jet are extremely consistent with values obtained from the previous data. The only difference is seen in the \xe\ values $\leq$1\arcsec\ of the source, which show a moderate increase from 0.3 to 0.4. The increased angular resolution of the NFM data also allows us to compare our values with those obtained by \citet{Coffey2008} using HST/STIS spectra at the base of the red-shifted jet (+0\farcs3); we find values compatible with their measurements for all of the examined parameters.

The diagnostic estimates from the 2014 data in the blue-shifted lobe were reported only for an offset of -2\arcsec, due to the uncertain origin of the [\oi] emission close to the source. These values are also plotted in Fig. \ref{fig:flux_ratios_BE}, although we note that this position falls at the edge of the [\oi] emission in the NFM data. At -2\arcsec\ the NFM data shows higher values of \eden\ and \nh\ (1.01 x 10$^{4}$ \percm\ and 1.5 x 10$^{4}$ \percm, respectively) with lower \xe\ = 0.45 and \Te\ = 3 x 10$^{4}$ K. 

We obtain directly comparable estimates of the gas properties within 1\arcsec\ on each side. At 0\farcs4, \eden\ and \xe\ are very similar between the two sides, but \Te\ is significantly higher on the blue-shifted side. At larger offsets the red-shifted lobe rapidly shows asymmetric properties with a sharp decrease in estimated \eden\ and \nh\ compared with a more gradual decrease in the blue-shifted lobe. Within 1\arcsec\ the density \nh\ in the two lobes is relatively similar, but beyond this region the density in the blue-shifted lobe becomes more than a factor 2 higher.

Both lobes show jumps in \xe\ at the estimated knot positions, however these are slightly larger in the blue-shifted jet, resulting in moderately higher \xe\ overall. \Te\ remains constant in the red-shifted jet beyond 1\arcsec\ but increases along the blue-shifted jet (with significant scatter in the estimated values). 

\subsection{Line ratios using [\oii]}
\label{subsection:oii_ratios}

As an additional check on the jet temperatures and electron densities, we examine four line flux ratios of [\oi] and [\sii] with [\oii], which have recently been used by \citet{Sperling2025} to explore these properties in jets. The [\oii] 7320\AA\ and 7330\AA\ doublets are each observed within both Th 28 jet lobes within 1-2\arcsec\ of the source, although the individual lines within each doublet are blended. We explore the ratios of these lines with the \texttt{pyneb} package \citep{Luridiana2015}. The [\oii] energy levels \citep{Martin1993}, A-values \citep{Zeippen1982} and collision strengths \citep{Kisielius2009} were obtained from the NIST atomic database.

\begin{figure*}
	\centering
    \includegraphics[width=15cm, trim= 0cm 0.cm 0cm 0cm, clip=true]{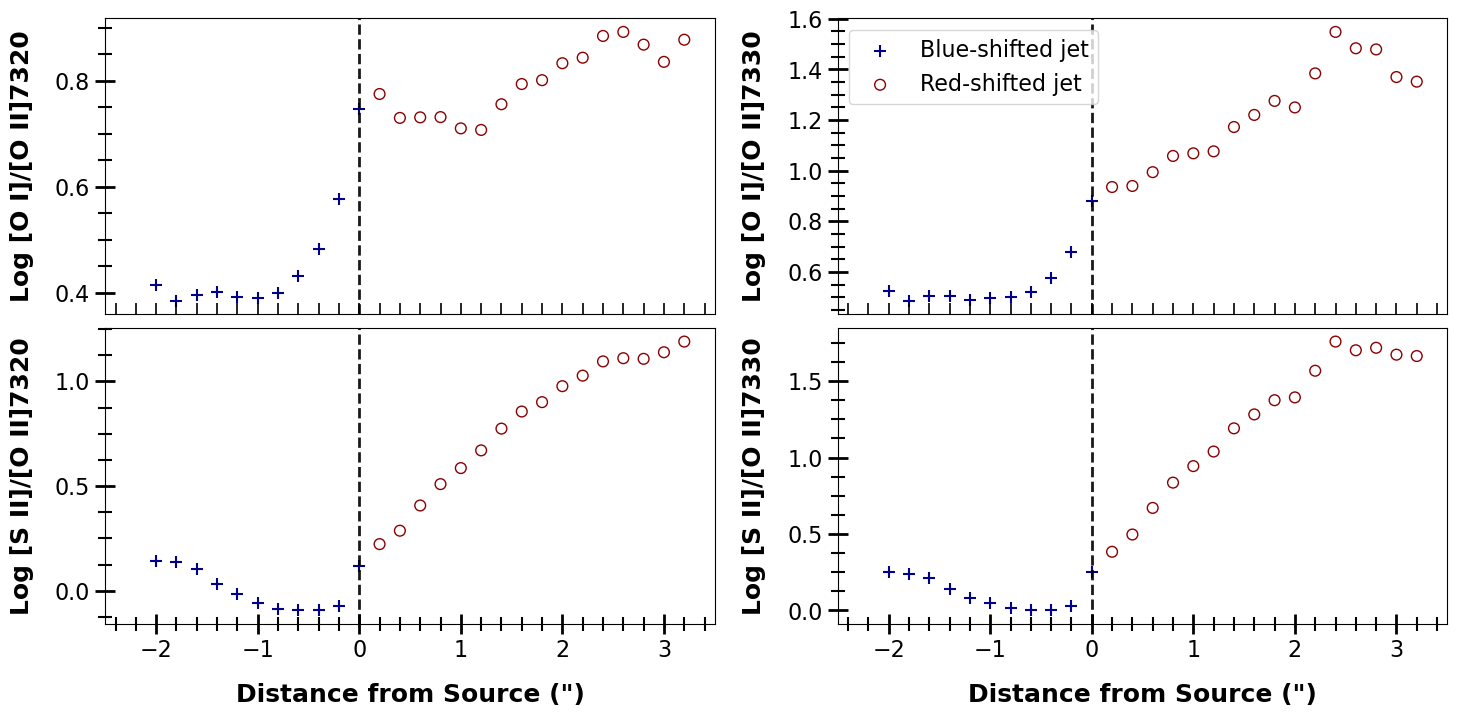}
	\caption{\label{fig:flux_ratios_oii}Flux ratios for emission lines of [\sii]$\lambda$(6716+6731) and [\oi]$\lambda\lambda$(6300+6363) with the [\oii] $\lambda$7320 \AA\ and $\lambda$7330 \AA\ lines, with fluxes sampled from the inner +/- 0\farcs2 of the jet axis. The jet axis is oriented as in Fig. \ref{fig:flux_ratios_BE}. All ratios are corrected for average \av\ = 1.0 mag.}
\end{figure*}

\begin{figure*}
	\centering
    \includegraphics[width=15cm, trim= 0cm 0.cm 0.3cm 0cm, clip=true]{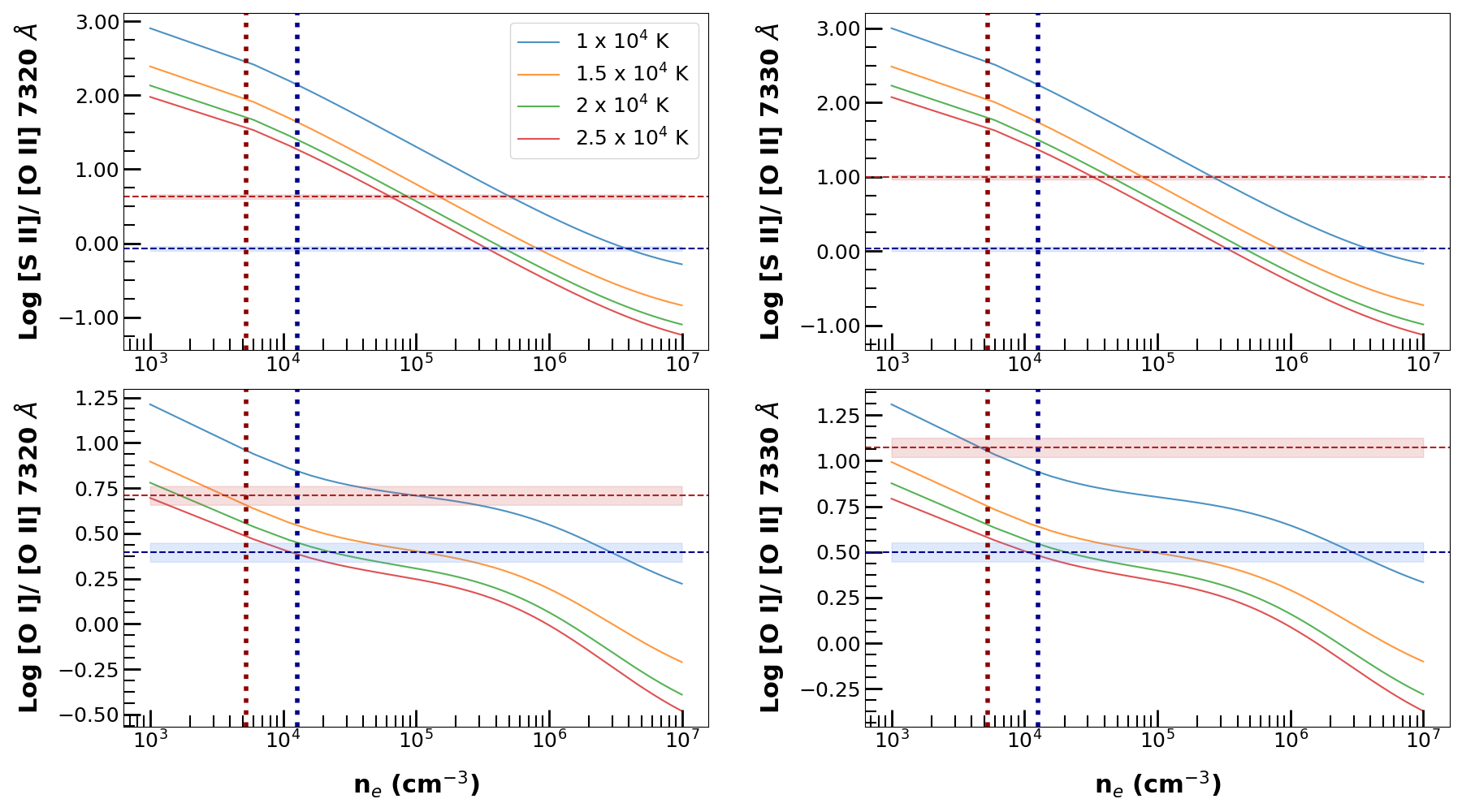}
	\caption{\label{fig:theory_oii} Theoretical curves for ratios of [\sii], [\oi] and [\oii]. Horizontal dashed lines show the average ratio values in the red- and blue-shifted jet lobes (red and blue lines), corrected for \av\ = 1 mag; the shaded regions show the minimum and maximum ratio values taking into account the uncertainty in flux calibration and allowing for a range of \av\ from 0 to 2 mag. Vertical dotted lines show the estimated \eden\ in the red- and blue-shifted lobes based on the BE method in Sect. \ref{subsection:be_ratios}.}
\end{figure*}

We examine the ratios [\sii]$\lambda$(6716+6731)/[\oii]$\lambda$7320, [\sii](6716+6731)/[\oii]$\lambda$7330, [\oi]$\lambda\lambda$(6300+6363)/[\oii]$\lambda$7320 and [\oi]$\lambda\lambda$(6300+6363)/[\oii]$\lambda$7330. Each of these ratios is inversely proportional to \eden, as well as showing a weak inverse dependency on \Te. The ratio values measured within the inner jet are shown in Fig. \ref{fig:flux_ratios_oii}. Since the wavelengths of the key lines differ by up to 1000 \AA, an extinction correction is applied assuming \av\ = 1.0 mag. Fig. \ref{fig:theory_oii} shows the theoretical curves for these ratios overplotted with the ratios measured at 1\arcsec\ from the source in the red- and blue-shifted jets. To account for the uncertainty in the extinction in both jet lobes (see Sect. \ref{subsection:extinction}), we include maximum and minimum values of these ratios for \av\ = 0-2 mag (also taking into account the MUSE flux calibration uncertainty of 5$\%$). These are compared with the \eden\ values measured at the same positions using the BE method in the previous section. 

In both [\oi] ratios, the measured values are consistent with the results of the BE method, with \eden\ of 6 $\times$ 10$^{3}$ \percm\ in the red-shifted jet and $>$1.3 $\times$ 10$^{4}$ \percm\ in the blue-shifted jet. They suggest a lower \Te\ of approximately 1-1.5 $\times$ 10$^{4}$ K in the red-shifted jet but agree with the estimated \Te\ of $>$ 2 $\times$ 10$^{4}$ K in the blue-shifted jet. Since these ratios are inversely proportional to temperature, the added [\oi] emission from the unresolved LVC may contribute to an apparently lower temperature.

The [\sii]/[\oii] ratios are less consistent with the BE results, indicating much higher \eden\ $>$ 10$^{5}$ \percm\ in both lobes. This may be due to the lower critical density of the [\sii] lines ($\sim$ 2 $\times$ 10$^{4}$ \percm), which means that these ratios are strongly suppressed close to the jet base where densities exceed this value. This would be consistent with the high values of \eden\ in the blue-shifted jet, where the [\sii]6716/6731 ratio is also seen to be in the high-density limit (Sect. \ref{subsection:be_ratios}). In the red-shifted lobe, the [\sii]6716/6731 ratio is not in the HDL; however, the [\oii] emission is seen from the inner knots, and may specifically originate in shock regions with higher ionisation and \eden\ than those traced by the BE method.

\subsection{Mass accretion rate}
\label{subsection:macc}

\begin{figure*}[h!]
\centering
\includegraphics[width=9cm, trim= 0.3cm 0.4cm 0.2cm 0cm, clip=true]{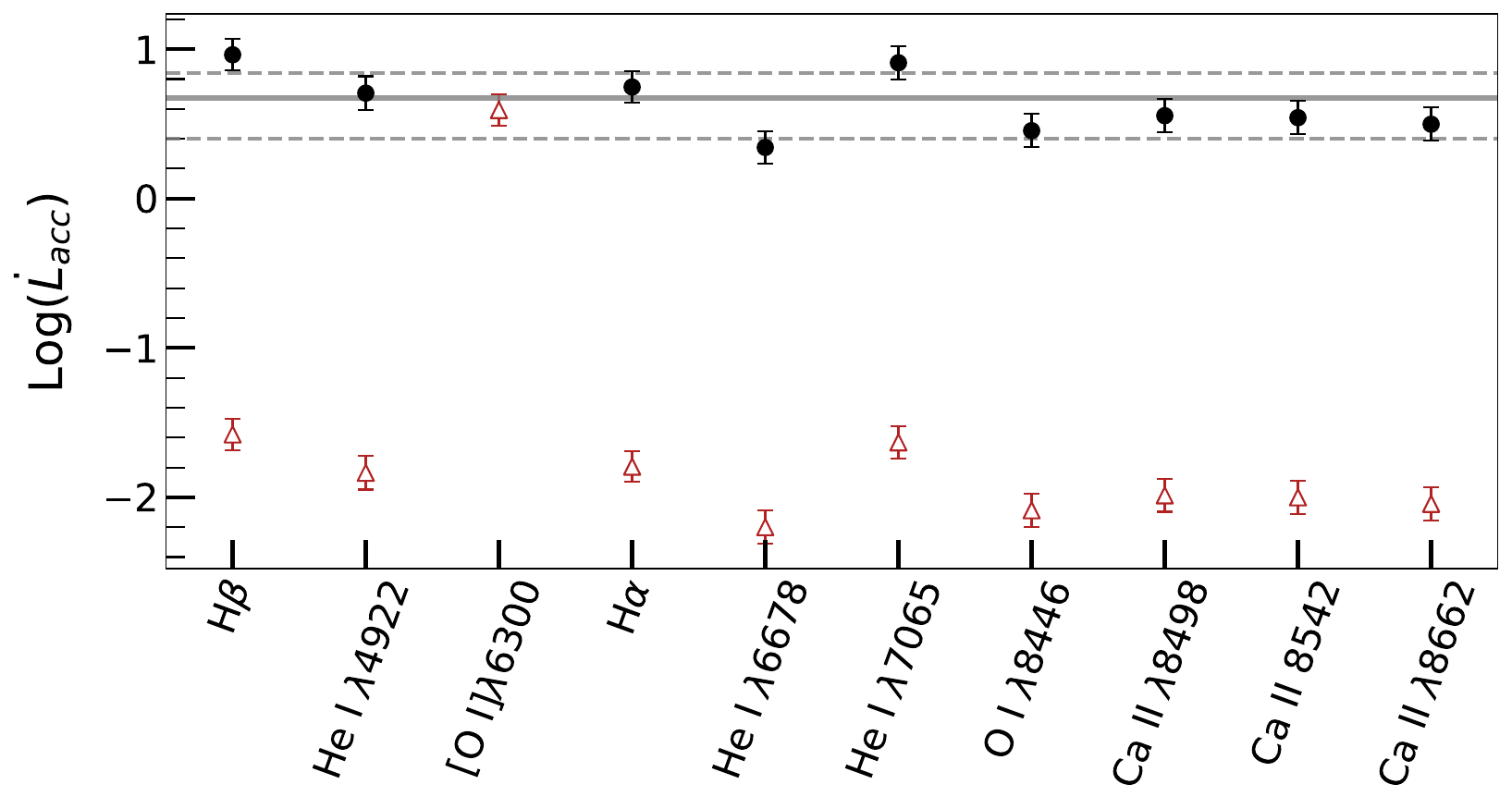}
\includegraphics[width=9cm, trim= 0.3cm 0.4cm 0.2cm 0cm, clip=true]{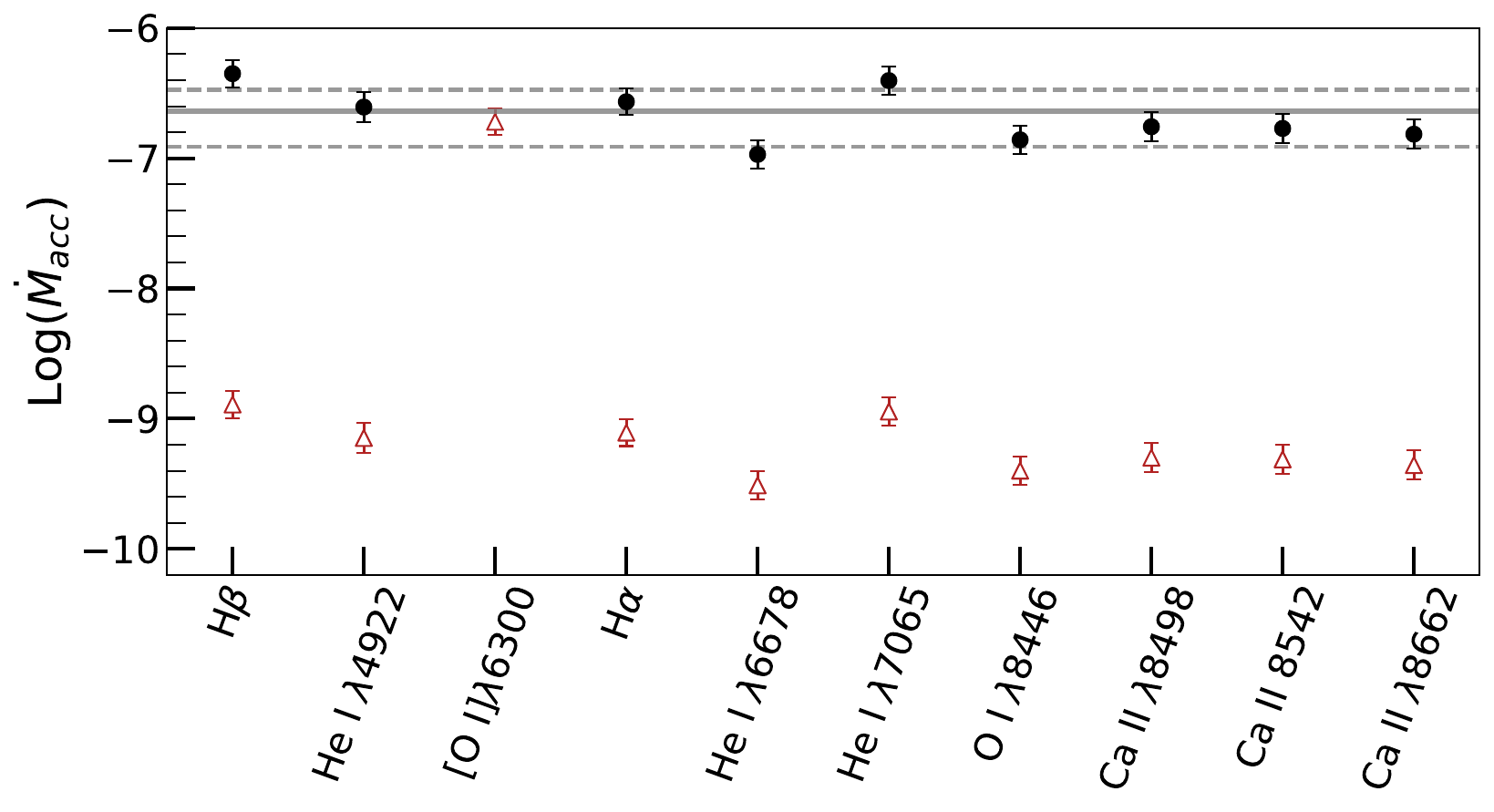}
\caption{\label{fig:macc_results} Estimated values of \Lacc\ (left) and \Macc\ (right), taking \av\ = 1.26 mag. Values are shown before (red) and after (black) correction for the estimated obscuration factor. In each plot the third datapoint shows the value derived from the \Ox\ line, which is used to estimate the correction for obscuration. The solid horizontal line shows the mean of the corrected values, with the 1-$\sigma$ bounds shown by dashed lines.}
\end{figure*}

We estimate the accretion luminosity \Lacc\ and hence the mass accretion rate \Macc\ from the spectrum at the source position, following the method described in \citet{Murphy2024}. This method is based on the empirically-determined correlations between the luminosity of accretion-tracing emission lines $L_{\mathrm{line}}$ and accretion luminosity \Lacc\ \citep{Herczeg2008, Alcala2014}. 

To check the region in which the source flux should be sampled, we selected several accretion tracers previously identified as having a minimal contribution from an extended jet component.  Fig. \ref{fig:macc_sampling} shows the line flux and estimated \Macc\ values obtained with increasing sampling radii around the central source position. We also extracted position-velocity (PV) maps of these lines to confirm the absence of significant jet emission and examine the spatial extension of the source emission (Fig. \ref{fig:macc_pvmaps}).

We find that the estimated \Macc\ plateaus at approximately 1\arcsec\ from the source, with $>$95$\%$ of the line flux being recovered within this region. From Fig. \ref{fig:flux_profile_models} we also note that this is the approximate region along the jet axis within which the \Ha\ flux is dominated by the source emission. For each accretion line we therefore extract the continuum-subtracted spectra within a box covering +/- 1\arcsec\ of the source. Each line is fitted with a Moffat peak and the integrated line flux obtained.

The edge-on disk of Th 28 substantially reduces the observed flux from the source position, resulting in both wavelength-dependent extinction and a wavelength-independent obscuration factor due to ‘grey scattering’. The estimated $L_{\mathrm{line}}$ from each spectrum was extinction-corrected for the averaged value of $A_{v}$ $\sim 1.26$ across the source region, as estimated from the [\feii] emission \citep{Murphy2024}. We estimate the obscuration factor using the \Ox\ line, which is also correlated with \Macc\ but which originates from the jet above the dust plane and should therefore be less suppressed by obscuration.
 
We obtain initial estimates of \Macc\ for each of our accretion tracers using stellar parameters of T$_{eff}$ = 4900 K and  \Lstar\ = 2.0 \Lsun\ \citep{Alcala2017}, with \Mstar\ = 1.6 \Msun\ \citep{Louvet2016}. As in \citet{Murphy2024}, \Rstar\ is then $\sim$ 1.96 \Rsun\ based on these values. The obscuration factor was derived from the average ratio between the value of \Macc\ derived from the \Ox\ flux and each of the permitted lines. We then correct the values of \Macc\ obtained from the permitted lines using this obscuration factor.

The full results are given in Table \ref{table:Macc_MUSE}, with corrected \Macc\ values given for the permitted lines. Fig. \ref{fig:macc_results} shows \Lacc\ and \Macc\ for the different lines, with the permitted lines shown before and after correction for obscuration. We obtain an estimate of \Macc\ = 1.25 $\times~10^{-7}$ \Msun~\peryr\ from the [\oi] line luminosity and an average obscuration of 61. Combining this with the corrected values of \Macc\ using our obscuration factor, we obtain an average \Macc\ = 2.11 $\times~10^{-7}$ \Msun~\peryr. 

\subsection{Mass outflow rates}
\label{subsection:mout}

\begin{figure}
	\centering
	\includegraphics[width=9cm, trim= 0.4cm 0cm 0cm 0cm, clip=true]{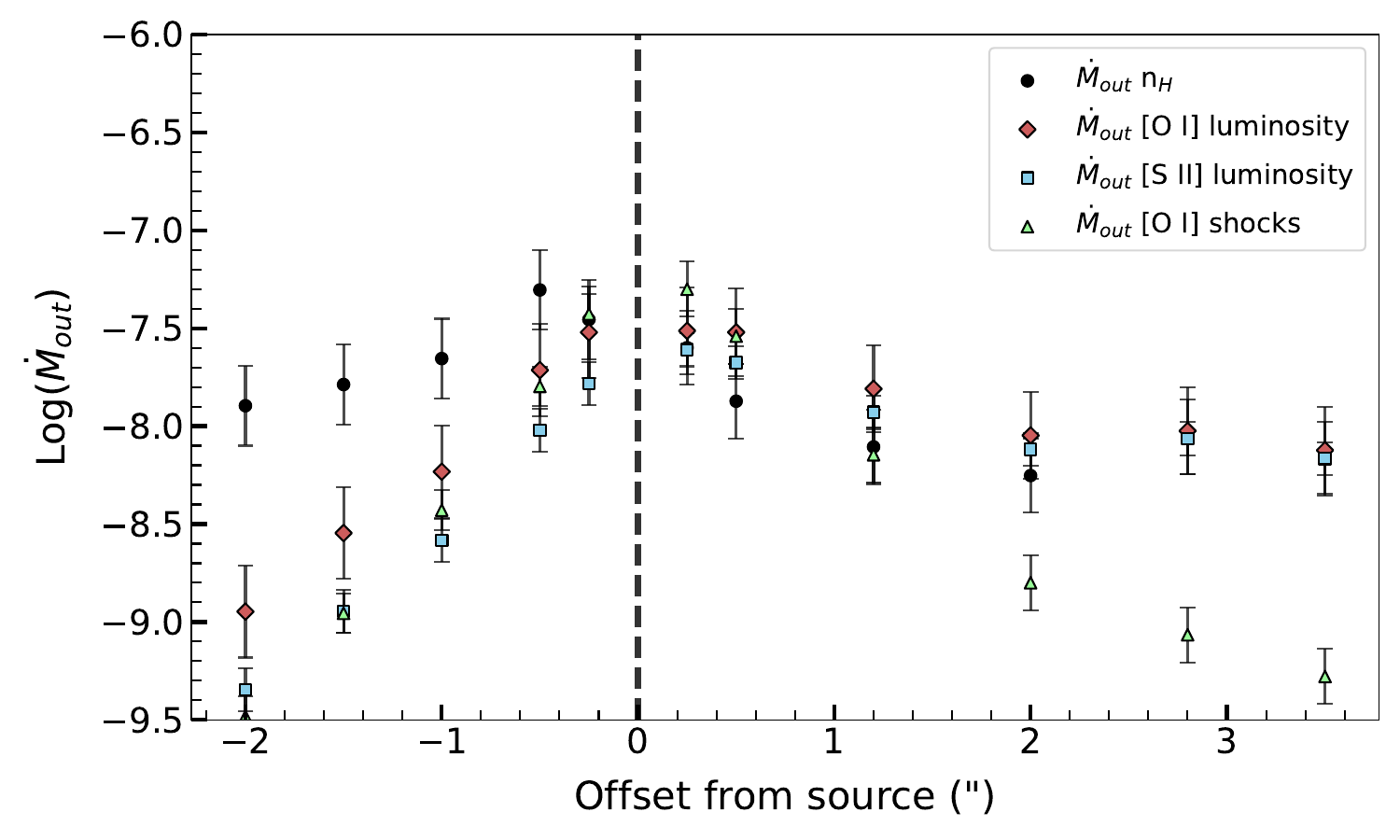}
	\vspace{-0.5cm}
	\caption{\label{fig:mout_rates}Mass outflow rates for the blue- and red-shifted jet (at negative and positive offsets respectively) derived using the three different methods denoted: \nh; luminosity; shock velocity. Estimated uncertainties are 20-50$\%$. The vertical dashed line marks the source position.}
\end{figure}

To estimate the rate of mass outflow (\Mout) through the jet we compare values obtained using three methods. As in the preceding paper, we obtain estimates by 1), combining the estimated \nh, jet velocities and jet cross-section in each lobe  ; 2), using the line luminosities of \Ox\ and \siib\ \citep{Hartigan1994, Bacciotti2011}; and 3), the method described by \citet{Agra-Amboage2009} which makes use of the \Ox\ line emission assuming that this originates from shock fronts within the aperture. We denote the measured outflow rates obtained from each of these methods as \Moutnh, \MoutLo,  \MoutLsii\ and \Moutshock, respectively.

The first method assumes a uniform density within the cross-sectional volume of the jet (i.e., a filling factor = 1). We use the estimates of the deconvolved jet FWHM obtained in Sect. \ref{subsection:jet_width}, and deprojected jet velocities of $v_{\mathrm{j,red}}$ = 270 \kms\ and $v_{\mathrm{j,blue}}$ = 365 \kms. These are combined with values of \nh\ estimated at the same jet positions using the unconvolved jet images to obtain \Moutnh. The second method using the FEL luminosities assumes uniform gas conditions and volume emission over the sampled aperture. It takes into account the electron density and ionisation fraction of the emitting gas, which are also obtained with the BE method at the sampled positions along the jet. Finally, we compare this with the third method which interprets the \Ox\ emission as originating from shock fronts. In the blue-shifted jet, we assume \vshock\ of 90 \kms, while for the red-shifted jet we take \vshock\ of 90-70 \kms\ at the knot positions $<$1.5\arcsec\ (since [\oiii] emission is associated with the inner 1\arcsec) and a lower limit \vshock\ of 30 \kms\ at larger offsets, following the line ratios in Sect. \ref{subsection:vshock}. We assume each aperture contains one line-emitting shock front. 

When sampling the FEL line emission, we wish to take into account the more diffuse emission of the blue-shifted jet. We thus examined the estimated line flux with increasing radius of the sampled region around the jet axis in each lobe. Fig. \ref{fig:jet_flux_radii} shows that even at an offset along the jet axis of +/- 0\farcs5, the collected flux in the red-shifted lobe flattens at smaller radii than in the blue-shifted lobe. In the red-shifted lobe, we therefore sample the emission within radii of +/- 0\farcs6 around the jet axis, whereas we extend this radius to +/- 1\arcsec\ in the blue-shifted jet. We do not apply a correction for extinction following the method in the previous paper, due to the low estimated extinction in both jet lobes.

Our estimated uncertainties in \Mout\ take into account uncertainties in the jet width, velocity, line flux and gas properties. For the shock front emission, the primary uncertainties are in the jet velocity and line flux. We adopt uncertainties of 10$\%$ and 20$\%$ on \vjetr\ and \vjetb, respectively, with uncertainties of 20$\%$ on the jet width. For the flux uncertainty we take the MUSE flux calibration uncertainty of 5$\%$, and from the diagnostic analysis we estimate uncertainties in \eden\ and \xe\ of 40$\%$ and 20$\%$, respectively. We take the uncertainty on \vshock\ to be 30$\%$ (based on the range of possible \vshock\ associated with the measured line ratios). Combining these uncertainties, we find approximate uncertainties of 50$\%$ on \Moutnh\ and \MoutLo, and 20-30$\%$ in \MoutLsii\ and \Moutshock. This is largely due to the dependence of the former values on \eden, which has the largest proportional uncertainty.

Fig. \ref{fig:mout_rates} shows the estimated \Mout\ values along the jet. In the red-shifted jet, \Mout\ ranges from 1-3 $\times$ 10$^{-8}$ \Msun~ \peryr\ at the jet base and flattens to approximately 0.5-2 $\times$ 10$^{-8}$ \Msun~ \peryr\  along the jet, with a small increase at the knot position around +3\arcsec. The lowest values are observed in \Moutnh, and this effect is most pronounced within 1\arcsec\ of the source, where a contributing factor may be additional [\oi] and [\sii] emission from the LVC combined with very narrow measured jet widths. However, this does not explain the lower values of \Mout\ using this method at larger offsets, which was not observed in the previous data. This may be due to the extended flux sampling used in this work, whereas the previous analysis used line luminosities sampled over a region of 1\arcsec\ $\times$ 1\arcsec\ to match the seeing-limited resolution. We also note that the values of \Moutshock\ are consistent with the other measurements close to the jet base (up to the first knot at 1\farcs2), but then drop sharply in contrast to the other methods. A similar pattern was observed by \citet{Agra-Amboage2009} when using this method to estimate \Mout\ in RY Tau; they observe that for a constant shock velocity along the jet, the estimated \Moutshock\ tends to decrease in accordance with the \Ox\ luminosity. This may be due to the lateral widening of the jet with distance, resulting in a decrease in \eden\ which this method does not compensate for. 

In the blue-shifted jet, the estimated \Mout\ from all methods converges at 1-3 $\times$ 10$^{-8}$ \Msun~ \peryr\ near the jet base. In this lobe, \Moutnh\ is consistently much larger than \MoutLo\ and \MoutLsii, as found in \citet{Murphy2024} This conforms with the higher jet velocity and \nh\ estimated in this lobe. Even with the more extensive flux sampling to compensate for the jet width, the estimates from [\oi] and [\sii] drop rapidly below 5 $\times$ 10$^{-9}$ \Msun~ \peryr. In this lobe, the estimate from \Moutshock\ remains in a similar range to those from line luminosity, with a less pronounced discrepancy along the jet axis.

\section{Discussion}
\label{section:discussion}

The previous WFM observations showed that Th 28 ejects knots on an estimated timescale of 10-15 years. With the increased angular resolution of the NFM-AO mode, we find a series of knots within the inner 6\arcsec\ of the red-shifted jet, outlined in Table \ref{table:knot_ids}. The possible knot at +2\arcsec\ is very tentative, and the knots at +2\farcs8 and +3\farcs5 may be either two merging structures or one marginally resolved knot centred at about +3\farcs2. 

Typical proper motions in the red-shifted jet have previously been well constrained at 0\farcs34 \peryr\ \citep{Comeron2010,Murphy2021}. We compare these with the knot proper motions measured between the MUSE WFM and NFM observations. The knot at +5\farcs9 may correspond to M4 from \citet{Murphy2021} (originally reported at 2\farcs5 and travelling 3\farcs4 in 9 years), while the knot at +4\farcs5 may correspond to the knot detected at +1\farcs2 by \citet{Melnikov2023} (with 8 years between epochs). In both cases, these would suggest unusually high proper motions of 0\farcs38 \peryr\ and 0\farcs41 \peryr, respectively. In the case of M4, this discrepancy may be due to uncertainty in the initial peak position, since it was not well resolved in the 2014 MUSE-WFM data. While the flux profiles indicate a knot peak at +2\farcs4, examination of the deconvolved spectro-images shows that the peak may be located at $\sim$+2\farcs8.  In this case, the estimated proper motion between epochs would again be 0\farcs34 \peryr. This explanation does not clearly apply to the other knot, which was detected using VLT/SINFONI observations with angular resolution $\sim$0\farcs7 and for which the peak can be seen in the [\feii] spectro-images. Therefore the higher proper motions in this knot may represent a change in the underlying outflow velocity.

The proper motions of the blue-shifted jet are more poorly constrained, but are estimated to be approximately 0\farcs47 \peryr\ \citep{Murphy2021}. Adopting this value, the knots located at +1\farcs2 and -1\farcs2 would both have an estimated launching period 2-3 years prior to the observations. Taking a conservative interpretation of the knot structures (i.e., excluding the possible knot at +2\arcsec\ and assuming one knot at +3\farcs2) we find that knots are ejected on a timescale of 3-6 years, significantly more frequently than the previous data indicated. High-resolution imaging of a few other T Tauri jets including DG Tau A and RY Tau have shown knot ejections occurring on similar timescales of a few years or less \citep{Takami2023}. These observations demonstrate that Th 28 exhibits similarly frequent outflow variations, fitting the broader pattern of active jets showing a hierarchy of knot ejections on different timescales. The tentative knots in this data may represent more frequent small ejections occurring between the larger variations; that is, smaller variations in the outflow velocity leading to fainter shock emission. Jets which exhibit such frequent ejections on timescales of just a few years present ideal cases to test the link between episodic mass accretion and outflow variability.

The low jet inclination and the velocity resolution of MUSE mean that small variations in radial velocity between the knots cannot be discerned. Careful measurement of the knot proper motions in future studies or high resolution echelle spectra (e.g. with X-shooter) would be required to trace possible variation in the knots themselves over time (for example, as carried out for the DG Tau jet by \citet{Takami2023}). However, the line ratios accessible with MUSE allow us to explore the outflow velocity changes driving the underlying shocks. As discussed in Sect. \ref{subsection:vshock}, the blue-shifted jet consistently shows line ratios corresponding to \vshock\ in excess of 90 \kms. It is difficult to identify any variation in shock velocities from the line ratios alone as this is the upper end of the shock velocities effectively traced by these ratios.

An alternative is to compare the emission line profiles with the planar bow shock models of \citet{Hartigan1987}. These show that for certain low-excitation lines (e.g. \Ha) we can estimate the shock velocity from the full width at zero maximum (FWZM) of the line profile. A comparison of several emission line profiles at the innermost knots in each jet is shown in Fig. \ref{fig:knot_profiles}. Unlike in the red-shifted jet, the broad line profiles of the blue-shifted jet allow the line profile to be moderately resolved. The NFM profile of \Ha\ at -1\arcsec\ shows a FWZM of approximately 500 \kms. This is consistent with the similarly high value of $\sim$ 400 \kms\ seen in the higher resolution X-Shooter echelle spectra presented for the blue-shifted knot in \citet{Murphy2024}. If these velocities represent the underlying velocity changes driving these shocks, it implies that the outflow velocity varies by a value on the order of the outflow velocity itself. A more detailed analysis of the knot dynamics and spectra in this lobe is needed to constrain the proper motions near the source and clarify the properties of this jet.

The line ratios in the red-shifted jet present a more nuanced picture. At the first knot position (+2\farcs8), the \Ha/\Hb\ and  [\sii]/\Ha\ ratios are consistent with \vshock\ = 70-80 \kms. The [\oi]/\Ha\ ratio at this position suggests \vshock\ 40-55 \kms, although we note that since the line ratio is inversely related to \vshock, excess \Ox\ emission from the unresolved LVC may lower the apparent \vshock. A \vshock\ of 70-80 \kms\ is also consistent with the derived values of \Moutshock\ in Sect. \ref{subsection:mout}. However, as noted in Sect. \ref{subsection:mout}, the shock velocities derived for the extended jet beyond this point (40-55 \kms) result in estimates of \Moutshock\ an order of magnitude lower than those derived using other methods, which yield values extremely consistent with one another. 

\subsection{Mass accretion and outflow}
\label{subsection:discuss_macc_out}

We obtain \Macc\ = 2.11 $\times$ 10$^{-7}$ \peryr (mean log \Lacc = -1.9), which represents a increase of a factor $\sim$4 from the value of 5.5 $\times$ 10$^{-8}$ \peryr\ obtained from the 2014 WFM data for \av\ = 1.26 mag. However, the latter estimate was obtained from flux sampled over a smaller aperture (1\arcsec\ $\times$ 1\arcsec) than our present measurements. To ensure an accurate comparison, we therefore recalculated \Macc\ using flux sampled within the same region. From this we obtain \Macc\ = 1.33 $\times$ 10$^{-7}$ \peryr, corresponding to a factor 2 increase. This rise is also reflected in the underlying accretion luminosity (before correction for obscuration); with mean log \Lacc\ = -2 compared with -2.4 in 2014.

In the previous WFM observations, \Moutnh\ could not be estimated within the inner 2\arcsec\ of the red-shifted jet due to the unresolved jet width, and was not estimated within 1\farcs4 of the blue-shifted jet due to suspected contamination from the red-shifted emission. We therefore focus our comparison on the values of \Mout\ obtained using FEL luminosities in the red-shifted jet. Close to the base ($\sim$+0\farcs5), we find that \MoutLo\ increases from 1.7 $\times$ 10$^{-8}$ \Msun~\peryr\ to 3 $\times$ 10$^{-8}$ \Msun~\peryr\ between 2014 and 2023. Similarly, \MoutLsii\ increases from 8.4 $\times$ 10$^{-9}$ \Msun~\peryr\ to 2.1 $\times$ 10$^{-8}$ \Msun~\peryr. This matches the factor 2 increase in estimated \Macc, although we note the significant uncertainties of 20-50$\%$ on all of the estimated values. The ratio of \Mout/\Macc\ (i.e, the one-sided jet efficiency) thus remains $\sim$0.1, corresponding well to the theoretical predictions of MHD jet launching models \citep{Ferreira2006}.

As noted, the [\oi] line flux (and to a lesser extent the [\sii]) may include unresolved LVC emission; we assume any such contribution is similar in both the present data and the previous WFM observations. We also assume a constant outflow velocity along the jet, which may obscure some of the variations in \Mout. In the blue-shifted jet, we can compare \Mout\ at -1\farcs4 in both epochs. In contrast to the red-shifted lobe, we find a decrease in the mass outflow rate, which drops from 6.5 $\times$ 10$^{-9}$ \Msun~\peryr\ to 2.7 $\times$ 10$^{-9}$ \Msun~\peryr\ for \MoutLo\ and 1.8 $\times$ 10$^{-9}$ \Msun~\peryr\ to 1.1 $\times$ 10$^{-9}$ \Msun~\peryr\ in \MoutLsii. In both epochs an unresolved knot is thought to be located at $\sim$-1\arcsec, but the positions of the knot peaks are poorly defined. Sampling the knot position in the 2023 data, we find \MoutLo\ =  5.4 $\times$ 10$^{-9}$ \Msun\ \peryr\ and \MoutLsii\ = 2.4 $\times$ 10$^{-9}$ \Msun\ \peryr, which are similar to the 2014 measurements. 

One consideration is the differing spatial resolution of the observations, since the aperture at -1\farcs4 in the 2014 data likely included emission from the nearby knot and may represent a similar heightened ejection phase. We also note the uncertain origin of the [\oi] emission in this lobe, which makes it unclear whether the [\oi] luminosity traces the jet mass loss; this also affects the estimates of \Moutshock\ and \Moutnh, which depend on using the [\oi] line to estimate the jet density. On the other hand, the [\sii] emission clearly traces the jet and should be minimally affected by the estimates of \eden\ obtained from the BE method, as these are well above the critical density of the [\sii] line in the blue-shifted lobe. The \MoutLsii\ values also do not show a significant increase. The dominant factors are most likely the poor S/N of the emission at these distances, and the uncertainty of the knot positions relative to the sampling aperture. 

An additional consideration is that the emission at the jet base may not represent the value of \Mout\ corresponding to the accretion state observed in the same observation. If knots are launched from an extended radius in the disk, corresponding to MHD wind models, then a time delay can be expected between knot launching and the subsequent rise in mass accretion. From \citet{Takami2020} the time delay based on the stellar luminosity and disk radius is estimated as:

    \begin{equation}    
        t_{delay} = 1.4~\times~10^{3} \left(\dfrac{L_{*}}{L_{\odot}}\right)^{-1/8} \left( \dfrac{r}{1~\mathrm{au}}\right)^{5/4}~\mathrm{days} 
    \end{equation}

Hence for \Lstar\ = 2 \Msun, we find expected time delays of 70 days to 3.5 years for r$_{launch}$ = 0.1 – 1 au. A knot launched at 1 au, travelling at $\sim$ 0\farcs34 \peryr\ in the red-shifted jet, would then reach an offset of 1\farcs2 by the time of the corresponding rise in \Macc. Interestingly, this is the location of the closest detected knot in our images. Thus it is possible that \Mout\ measured in this knot corresponds to the observed accretion rate onto the star, and here we also find an increase of a factor 2 in \Mout\ using the FEL luminosities. 

On the other hand, the actual time delay between knot launching and accretion bursts could be significantly larger or smaller, and the current accretion burst may instead have been preceded by an as-yet undetected knot closer to the source. Thus, because the short timescales in the jet variability (hence underlying episodic accretion) are comparable to the possible delay between ejection and accretion, analysing the changes in \Mout\ relative to \Macc\ will require not only high spatial resolution to accurately track the knot positions, but careful time monitoring to match sampled values of \Mout\ to the corresponding values of \Macc\ \citep{Takami2020}. 

\subsection{Properties of the jet asymmetry}
\label{subsection:discuss_asym}

\begin{figure}
\centering
\includegraphics[width=8.5cm, trim= 0cm 0cm 0cm 0cm, clip=true]{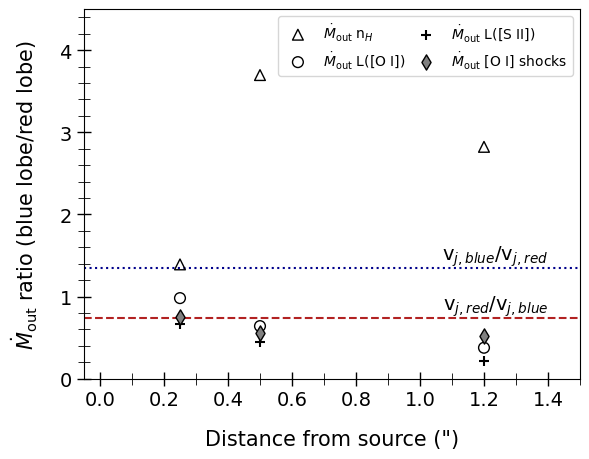}

\caption{\label{fig:mout_asymmetry}The ratio of \Mout\ measured in the blue-shifted jet versus the red-shifted jet, plotted against distance from the source. The red dashed line indicates the the ratios \vjetr/\vjetb, corresponding to the case where linear momentum is conserved; the blue dashed line indicates \vjetb/\vjetb, the case in which the mass outflow rate is larger in proportion to the jet velocity}
\end{figure}

An overview of the jet properties in each lobe is given in Table \ref{table:asymmetries]}. Previous observations have shown the contrasting morphology of the extended jets. However, the sub-arcsecond angular resolution of the NFM data allows us to resolve the shape of the compact red-shifted knots for the first time. The knots within 3-6\arcsec\ exhibit arc-shaped morphologies resembling bow shocks, suggesting that the knots closer to the source may similarly be unresolved bow shocks. 

As discussed in Sect. \ref{subsection:jet_width}, the jet width measurements in each lobe may be strongly affected by the extent of the shock wings. Both lobes show a similar opening angle (20-25\ndeg) when measured in \Ha\ (the brightest line in the blue-shifted lobe), as well as the FEL emission for the red-shifted jet. The initial flow collimation in the two lobes may therefore be more similar than previously estimated, with the expansion of the faster, broader shocks in the blue-shifted jet giving rise to the apparent asymmetry in collimation found in the MUSE-WFM observations \citep{Murphy2021}. We note that the opening angle of the blue-shifted lobe is substantially narrower (7-13\ndeg) in [\nii] and [\sii], which primarily trace the centre of the outflow. The discrepancy between FELs and \Ha\ in this lobe may reflect the diffuseness of the emission due to the broad bow shocks, resulting in much poorer S/N for the FELs emitted from the knot centre.

The FEL diagnostics employed by the BE method suggest that close to the base ($\leq$ 0\farcs5) the \eden, \xe, \Te\ and \nh\ are similar in each side of the jet. In both cases \eden\ and \nh\ decrease rapidly with distance from the source while \xe\ increases (Fig. \ref{fig:flux_ratios_BE}). However, by the first knot position ( $\sim$1\arcsec\ in both lobes) there is a pronounced difference with the blue-shifted jet showing the markedly higher values of \Te, \xe\ and \nh\ estimated at larger offsets. This is consistent with the gas excitation being driven by the internal shocks along the outflow, and thus heightened by the faster shocks in the blue-shifted jet. These are again highlighted by the shock tracing line ratios examined in Sect. \ref{subsection:vshock}, which indicate \vshock\ $<$ 70 \kms\ in the red-shifted lobe in contrast with the high-velocity shocks ($>>$ 90 \kms) indicated by both line ratios and emission line profiles in the blue-shifted lobe.

We caution that the main contribution to [\oi] emission in the blue-shifted jet remains unclear. Previous X-Shooter spectra indicated the presence of a narrow LV peak (\vrad\ = $<$ 10 \kms) which extended along the blue-shifted jet; this is seen again here in the blue-shifted jet line profiles, extending to offsets of $>$2\arcsec. This may represent a narrow LVC component tracing the interaction region between the jet and wind rather than the HV jet itself. However, inspection of the cube reveals diffuse [\oi] emission with a similarly narrow velocity profile around the source which is not confined to the jet axis. This effect is much less pronounced in the [\sii] emission and [\nii] lines which are also bright in the red-shifted jet, making it unlikely that this diffuse emission solely represents reflected emission from this lobe. Given the high excitation and emission of [\oii] and [\oiii] in the blue-shifted lobe, the jet outflow may contain very little neutral O; therefore a more detailed diagnostic which utilises these higher excitation lines is necessary to converge on accurate properties for the blue-shifted jet, in particular the ionisation fraction.

As a first step towards this, we thus explored additional diagnostics utilising [\oii] emission lines in Sect. \ref{subsection:oii_ratios}. Long-slit VLT/UVES and X-Shooter spectra of the Th 28 jet have identified the presence of several additional lines tracing regions of high electron density within the inner 1-2\arcsec\ of both lobes, including [\oii]$\lambda\lambda$3726, 3729 and [\sii]$\lambda\lambda$4069, 4076 \citep{Liu2014,Liu2021, Murphy2024}. \citet{Liu2014} made use of the [\sii] $\lambda$4069/\siib\ ratio to carry out diagnostics of the red- and blue-shifted emission, obtaining \eden\ = 6.6 $\times$ 10$^{4}$ \percm\  and 4.6 $\times$ 10$^{4}$ \percm\ in the red- and blue-shifted jets, respectively. They correspondingly estimate higher \xe\ in the blue-shifted jet by a factor of 2-3, with lower \nh\ in the red-shifted jet (we note that values of \xe\ and \nh\ were not reported). While these results are hampered by emission blending due to the seeing-limited angular resolution of their data (1\farcs5), a less dense blue-shifted jet is more consistent with the higher excitation in this lobe. Examination of the UV [\oii] lines in the jet also supports this picture, with \eden $>$ 10$^{5}$ \percm \citep{Liu2014, Liu2021}.  

Furthermore, Th 28 is also one of just three jets in which the [\neiii]$\lambda\lambda$3869, 3967 emission lines have been detected, which are sensitive to \eden\ $\leq$ 10$^{7}$ \percm\ in high-excitation regions \citep{Liu2021}; however, no diagnostic study of the jet has yet been carried out with these lines. These findings reinforce not just the importance of utilising emission lines tracing a wide array of excitation conditions to obtain an accurate picture of the gas properties, but the availability of these tracers in the Th 28 jet. Combining multi-wavelength diagnostics is necessary to obtain accurate estimates of the jet density and hence \Mout\ on each side of the jet. 

The ratio between the mass outflow rates in each lobe is of particular interest to understand the connection between mass loss and the asymmetry of the jet. Fig. \ref{fig:mout_asymmetry} shows the ratio between the blue- and red-shifted mass outflow rates. We find that close to the jet base, the mass outflow rates in each lobe converge to comparable values using all methods. If the outflow rate balances linear momentum, this ratio should fall close to the ratio of \vjetr/\vjetb. Indeed, at small offsets the outflow ratio measured using FEL emission falls closest to this value. In contrast, the ratios of \Moutnh\ are much larger due to the high values estimated for this method in the blue-shifted jet, which may not be reliable due to their dependence on the diagnostic estimate of \xe. In comparing the outflow rates in each lobe, we must also apply the caveats discussed in the previous section regarding the [\oi] line emission in the blue-shifted jet and the applicability of the BE diagnostics in this region.  

\begin{table}
\centering
\caption[target]{Summary of the properties in the red- and blue-shifted jet lobes.}
\vspace{0.1cm}
\renewcommand{\arraystretch}{1.4}
\begin{tabular}{{p{0.16\textwidth}<{\raggedright} p{0.12\textwidth}<{\raggedright} p{0.11\textwidth}<{\raggedright} }}
\hline \hline
Property & Red-shifted jet & Blue-shifted jet\\
\hline

$\theta_{\mathrm{H}\alpha}$ (\ndeg) & 21 & 25 \\
$\theta_{[\sii]}$ (\ndeg)& 20 & 6-8\\
PM (\arcsec~\peryr) & 0.34\footnote[1]{1} & 0.47\footnote[1]{1} \\
\vtan\ (\kms) & 266 & 340 \\
\vrad\ (\kms) & +20 & -25 to - 70\\
\vjet\ (\kms) & 270 & 345\\
$i$ (\ndeg)  & +4-5 & -4 to -12\\
\vshock\ (\kms) & 40-70\footnote[2]{2} & $>$90\footnote[3]{3} \\
 \multicolumn{3}{l}{\vspace{-0.2cm} } \\
\eden\ (\percm) & 1.3 $\times$ 10$^{4}$ & 1.47  $\times$ 10$^{4}$\\
\xe\  & 0.25 & 0.21\\
\Te\ (10$^{4}$ K) & 1.9 & 2.7\\
\nh\ (\percm) & 5.7 $\times$ 10$^{4}$ & 7.7  $\times$ 10$^{4}$\\

 \multicolumn{3}{l}{\vspace{-0.2cm} } \\
\Mout\footnote[4]{4} (10$^{-8}$ \Msun\ \peryr) &  & \\
~~~~  \nh\ method & 13.4 \small($\pm$ 5.8) &   \\
~~~~  $L_{\mathrm{[\oi]}}$ method & 30.1 ($\pm$ 15.4) & 19.3 ($\pm$ 10.6)  \\
~~~~  $L_{\mathrm{[\sii]}}$ method & 21.1 ($\pm$ 4.1) & 9.5 ($\pm$ 2.4)  \\
~~~~  Shocks method & 28.8 ($\pm$ 9.3) & 16.0 ($\pm$ 3.7)  \\
 \multicolumn{3}{l}{\vspace{-0.2cm} } \\
\hline
\multicolumn{3}{l}{\vspace{-0.4cm} } \\
\multicolumn{3}{l}{ \makecell[l]{Notes: (1) From \citet{Murphy2021}. (2) Between +0\farcs5 \\ and +3\farcs5. (3) Between -0\farcs5 and -2\arcsec. (4) \Mout\ sampled at \\ 0\farcs5 separation  from the source position.}} \\
\multicolumn{3}{l}{\vspace{-0.4cm} } \\
\hline
\end{tabular} 

\label{table:asymmetries]}
\end{table}

\section{Conclusions}
\label{section:conclusions}
The NFM-AO observations of the Th 28 jet enable the emission from the two jet lobes to be resolved within 200 au of the source, allowing a spatial and spectral comparison of their properties close to the base. This data also reveals the detailed structure and outflow variability of the inner 6\arcsec\ of the red-shifted jet lobe for the first time. Our main conclusions are therefore as follows:
\begin{enumerate}

\item{The jet widths are resolved to within 0\farcs5 (80 au) of the source, and opening angles of 5-11\ndeg\ are measured in the red-shifted jet. The blue-shifted lobe is likely strongly affected by unresolved broad bow shocks, exhibiting a sharp discrepancy between the opening angles of 6-8\ndeg\ in [\sii] and 25\ndeg\ in \Ha. This is supported by the complex kinematics in this lobe as demonstrated by the double-peaked line profiles and evidence of high-velocity shocks (\vshock\ $>>$ 90 \kms). }

\item{The deconvolved spectro-images reveal a newly ejected knot at $\sim$1\arcsec\ in the blue-shifted jet and 2-4 new knots within 3\arcsec\ of the red-shifted jet. Two knots previously identified in the red-shifted jet are also detected allowing their proper motions to be inferred. The most recent detectable knots were ejected 2-3 years ago and the separation between knots suggests episodic ejection on timescales of 3-6 years. It is thus likely that new knots have been ejected from both lobes in the 2 years since these observations.}

\item{We obtain \Macc\ = 2.11 $\times$ 10$^{-7}$ \Msun\ \peryr, corresponding to a factor 2 increase between 2014 and 2023. Comparison of the mass outflow rates in the red-shifted jet shows a matching increase in the outflow rate, with average \Mout\ = 1-3 $\times$ 10$^{-8}$ \Msun\ \peryr.}

\item{We measure the mass outflow rates in both lobes close to the jet base for the first time. At offsets $<$ 1\arcsec\ we find that values of \Mout\ converge to average values of 1-3 $\times$ 10$^{-8}$ \Msun\ \peryr\ in both lobes. This yields an overall jet efficiency of 0.1, in good agreement with MHD jet launching models. However, within these averages there is significant nuance between results obtained using different indicators. The ratio  in mass outflow between the two lobes corresponds best to a scenario where equal linear momentum is lost through both outflows, but more robust estimates of \Mout\ in the blue-shifted lobe are needed to confirm this.}
\end{enumerate}

These observations highlight the active accretion and corresponding outflow variability in the Th 28 system, comparable to the short-term variable ejection seen in a handful of other CTTS jets. This makes the Th 28 jet an ideal target for future monitoring observations at high spatial resolution to further investigate the connection between mass accretion and ejection. In particular, because of the symmetric ejection patterns observed in both jet lobes, Th 28 may be an excellent candidate to test the relationship between episodic ejection and the asymmetric jet properties. Future work is needed, however, to accurately constrain the mass outflow rate and proper motions within the highly ionised blue-shifted jet. Extended diagnostic methods which exploit the high-ionisation emission lines exhibited in this jet offer a promising avenue to achieve this.  

\section{Data Availability}
The Phase 3 observational data for this project can be accessed through the ESO Science Archive at \url{https://archive.eso.org/scienceportal/home?prog_id.keyword=110.23ZG}.

\begin{acknowledgements} 
We thank the anonymous reviewer for their comments and feedback. A. M. and M. T. acknowledge the grant from the National Science and Technology Council (NSTC) of Taiwan 112-2112-M-001-031-, 113-2112-M-001-009- and 114-2112-M-001-002-. M.B. is funded by the European Union (ERC, WANDA, 101039452). Views and opinions expressed are however those of the author(s) only and do not necessarily reflect those of the European Union or the European Research Council Executive Agency. Neither the European Union nor the granting authority can be held responsible for them.
\end{acknowledgements}	
\bibliographystyle{aa}
\bibliography{Bibliography}

\clearpage

\begin{appendix}

\section{Scattered emission subtraction}
\label{section:scattered_sub}

\begin{figure*}[ht!]
\centering
\includegraphics[width=14cm, trim=0cm 0.3cm 0cm 0cm, clip=true]{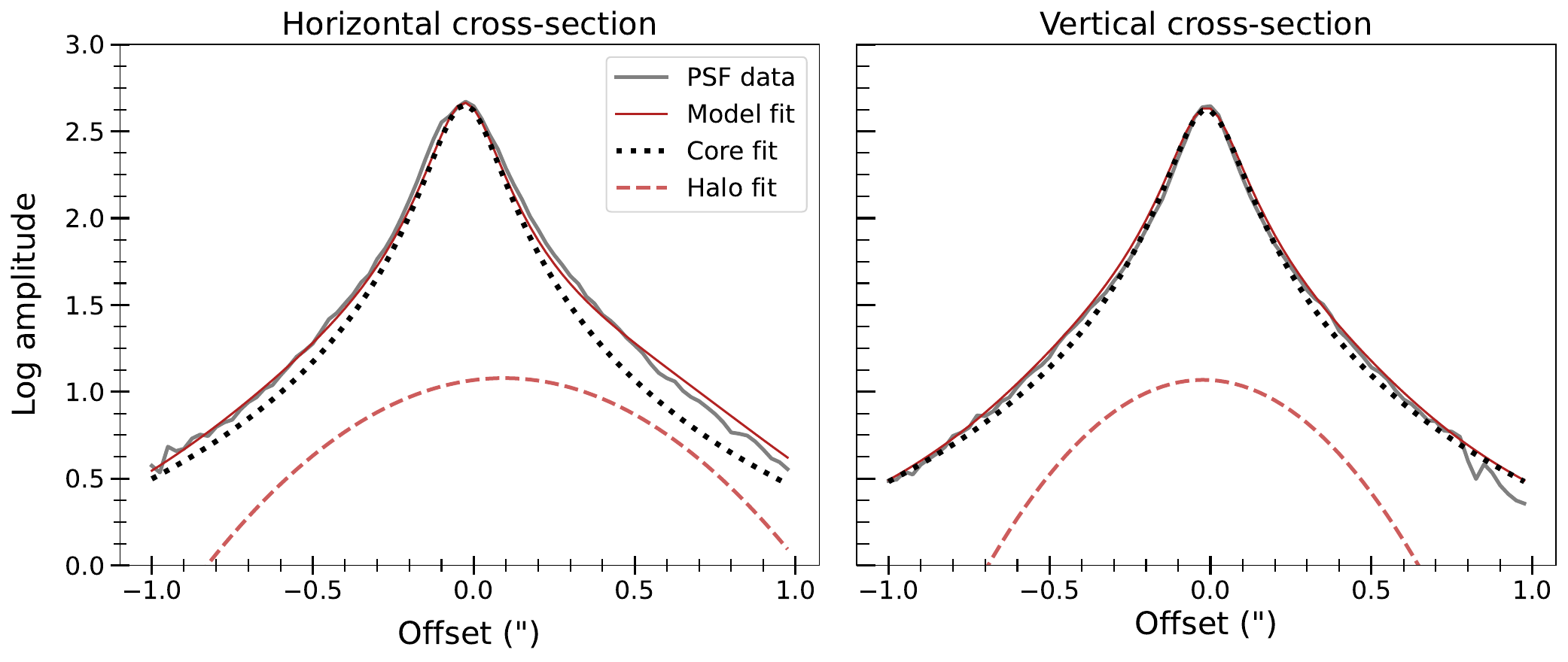}
\caption{\label{fig:continuum_model_cross_section}Cross-sections of a continuum model for reflected emission subtraction, fitted to a continuum image centred at 6500 \AA. The profiles are sampled parallel to the jet axis (horizontal, left panel) and perpendicular to the jet (vertical, right panel).}
\end{figure*}

A challenge in distinguishing the structure of the innermost Th 28 jet lobes (within +/- 1\arcsec\ from the source) is the significant diffuse line emission from around the source position. This occurs even in FELs which trace the jet primarily and may be due to emission scattered from the compact disk and envelope in this region. Therefore, in addition to the subtraction of the continuum emission described in Sect. \ref{section:target_obs} we also attempt to estimate and subtract this diffuse line emission in order to better highlight the jet morphology. 

For this procedure, we assume that the scattered jet emission will follow the same spatial distribution as the continuum emission; therefore at a given wavelength, the ratio between line and continuum emission should be constant in regions dominated by scattered emission. We can therefore model this distribution using the continuum emission datacubes obtained in Sect. \ref{section:target_obs}. 

For key emission lines, we then extract spectro-images of the jet and continuum over the same wavelength bins, and divide these to obtain line-to-continuum ratio maps. Two regions on either side of the source (perpendicular to the jet axis) were identified where the ratio between line and continuum emission was approximately flat (i.e., the spatial regions where scattered line emission dominates). From these regions we can then estimate the line-to-continuum (LTC) ratio in each wavelength bin, i.e. the LTC 'spectrum'. 

This spectrum is then scaled to the continuum level at each spaxel in the field of view. To avoid sharp discontinuities in the subtracted emission, we do not scale to the continuum image directly, but fit the continuum emission at each wavelength with a composite astropy model comprising a circular Moffat core and a wide Gaussian component, as illustrated in Fig. \ref{fig:continuum_model_cross_section}. When multiplied by the LTC spectrum, this results in a datacube of the estimated scattered emission, which can be subtracted from the line emission data. 

\begin{figure*}
\centering
\includegraphics[width=16 cm, trim=0cm 0cm 0cm 0cm, clip=true]{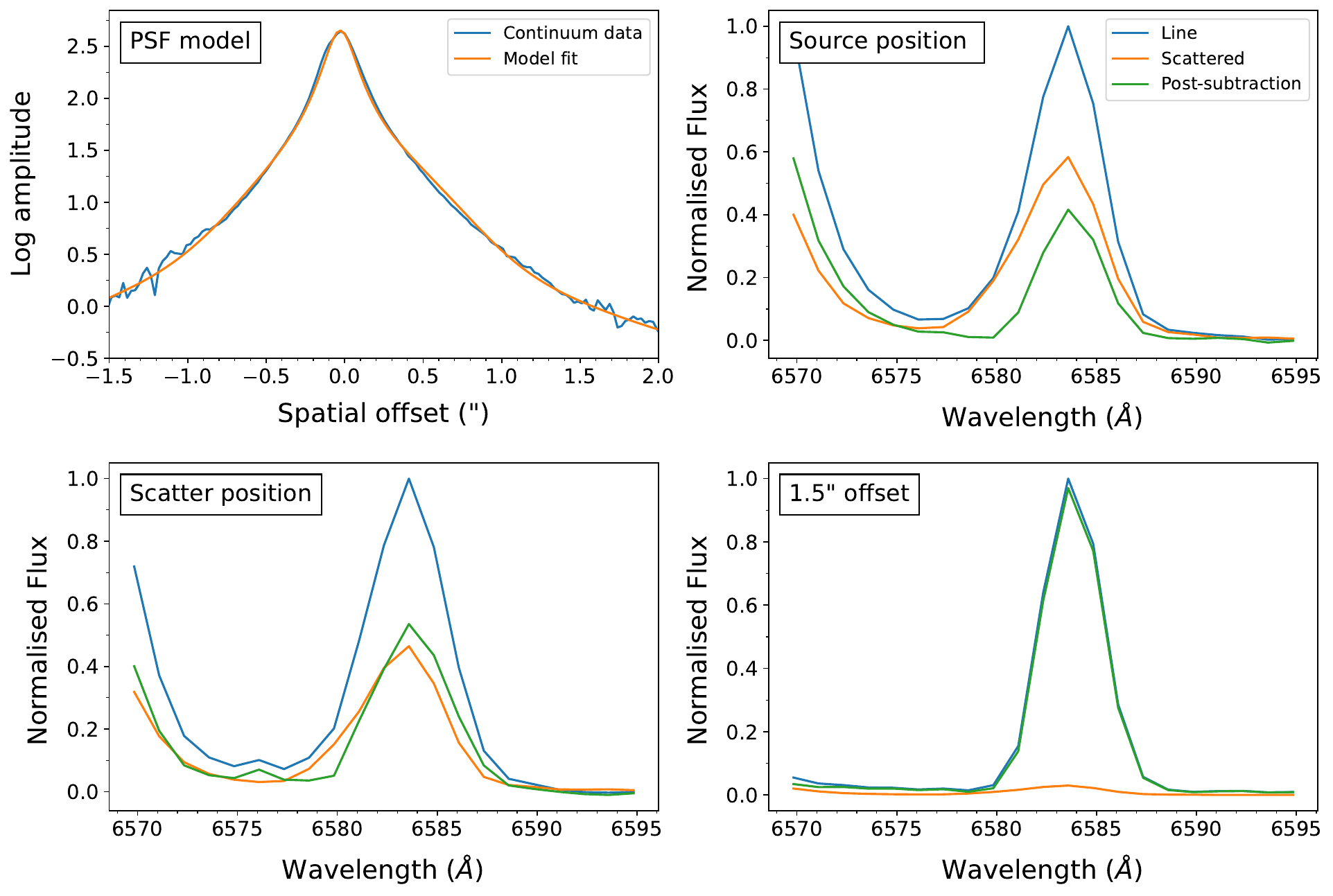}
\caption{\label{fig:nii_scattersub_report}Scattered emission subtraction report for the \niia\ line. Top left: cross-section of the fitted continuum model used to scale the subtracted spectrum, extracted parallel to the jet axis. Remaining panels show the line emission before and after subtraction of the scaled reflected spectrum, at the source position (top right); at one of the positions where the scattered emission is measured (bottom left) and at an off-source jet position (bottom right).}
\end{figure*}

To minimise over-subtraction of the jet emission, we introduce an additional scaling factor. We find that continuum emission drops to the background level at offsets of $\sim$ +/- 1\farcs5 in the jet axis direction. We therefore scale the subtracted line emission such that it falls to $\sim$0 at this distance, where the remaining reflected emission should be minimal. Fig. \ref{fig:nii_scattersub_report} shows an example of the subtracted spectrum at the source position, a scattering-dominated position and at +2\arcsec\ along the red-shifted jet. 

This procedure helps to reduce the strong line emission seen around the central +/- 0\farcs5-1\arcsec\ of the source, thereby highlighting the positions of the inner jet knots as shown in Fig. \ref{fig:flux_profiles}. However, it results in strong over-subtraction at the source position, where the continuum emission is no longer dominated by scattering; the scaled spectrum therefore becomes much higher than the observed line emission in this region. 

In addition, scaling the subtracted emission to fully remove the diffuse scattered emission around the source results in over-subtraction in the jet regions, particularly in the faint blue-shifted lobe, whereas avoiding over-subtraction using the scaling criteria described above results in minimal change to the observed spectro-images or line ratios. We conclude that it is difficult to accurately estimate the scattered jet emission separate to other components with differing spatial distributions, e.g. a wide-angled wind or scattered emission from a cavity wall.

In particular, this method is least effective for the \Ha\ and \OIa\ lines which are likely include contributions from non-jet components of other emission components (from accretion and from a wide-angled wind, respectively).We therefore cannot ensure consistency in the subtraction between different emission lines, such that the underlying flux ratios are preserved. Thus we only apply this procedure to explore the inner knot positions in Sect. \ref{subsection:knot_ids}, and do not use it when extracting the line fluxes used in the diagnostic analysis.

\clearpage
\onecolumn

\section{Knot widths and line profiles}
\label{section:knot_widths}
\begin{figure}[h!]
\centering
\includegraphics[width=18 cm, trim=0cm 0.4cm 0cm 0cm, clip=true]{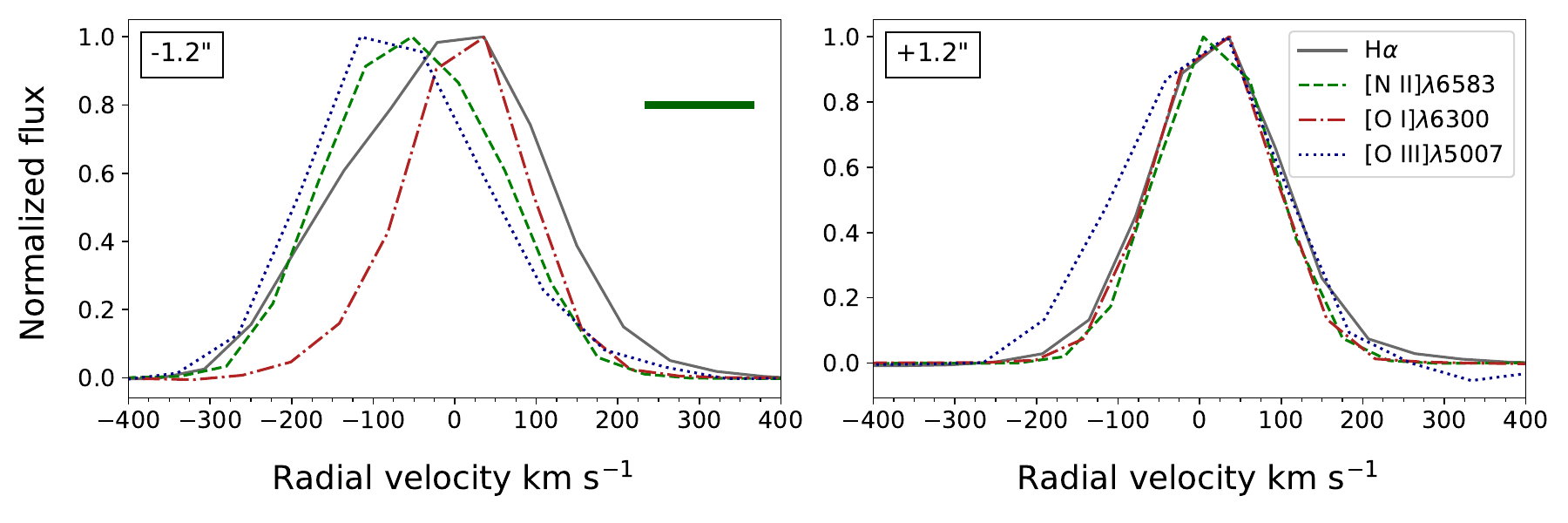}
\caption{\label{fig:knot_profiles}Comparison of emission line profiles sampled in the first blue- and red-shifted knots (left and right panels, respectively) from the MUSE-NFM data. The green bar indicates the approximate velocity resolution at wavelengths $\sim$6500 \AA (125 \kms).}
\end{figure}

\begin{center}
\begin{threeparttable}

\renewcommand{\arraystretch}{1.5}
\caption[target]{\label{table:knot_vrad}Peak radial velocity of the knot line profiles.}
   \begin{tabular}{{p{0.08\textwidth}<{\raggedright} p{0.08\textwidth}<{\raggedright} p{0.08\textwidth}<{\raggedright} p{0.08\textwidth}<{\raggedright}  p{0.08\textwidth}<{\raggedright} p{0.08\textwidth}<{\raggedright} p{0.07\textwidth}<{\raggedright}}}
    \hline \hline
    \makecell[l]{Offset \\ (\arcsec)}& \makecell[l]{\Ha \\ (\kms)}  & \makecell[l]{\siia\ \\ (\kms)} & \makecell[l]{\siib\ \\ (\kms)} & \makecell[l]{\niia\ \\ (\kms)} & \makecell[l]{\Ox\ \\ (\kms)} & \makecell[l]{\OIII\ \\ (\kms)}\\
    \hline
    -1.2 & -14.3 & -14.7 & -16.5 & -54.9 & 9.4 & -76.4 \\
    1.2 & 16.7 & 15.0 & 15.6 & 16.3 & 12.3 & -1.8 \\
    2.0 & 18.6 & 19.6 & 19.9 & 19.1 & 15.5 & 2.9 \\
    2.8 & 18.7 & 24.6 & 24.0 & 21.3 & 19.1 &  \\
    3.5 & 17.3 & 22.6 & 24.0 & 17.8 & 19.8 &  \\
    4.5 & 23.8 & 27.9 & 30.0 & 22.4 & 26.8 &  \\
    5.9 & 20.7 & 23.6 & 23.9 & 22.1 & 17.9 &  \\
    \hline
    \end{tabular}
    \begin{tablenotes}[flushleft]
	\small {
	\item[]Notes: All values are obtained from fitting a single Gaussian peak to the line profile.
    }
    \vspace{0.1cm}
    \end{tablenotes}
    \hrule
\end{threeparttable}
\end{center}

\begin{center}
\begin{threeparttable}

\renewcommand{\arraystretch}{1.5}
\caption[target]{\label{table:knot_fwhms}FWHM of the knot line profiles.}
   \begin{tabular}{{p{0.08\textwidth}<{\raggedright} p{0.08\textwidth}<{\raggedright} p{0.08\textwidth}<{\raggedright} p{0.08\textwidth}<{\raggedright}  p{0.08\textwidth}<{\raggedright} p{0.08\textwidth}<{\raggedright} p{0.07\textwidth}<{\raggedright}}}
    \hline \hline
    \makecell[l]{Offset \\ (\arcsec)}& \makecell[l]{\Ha \\ (\kms)}  & \makecell[l]{\siia\ \\ (\kms)} & \makecell[l]{\siib\ \\ (\kms)} & \makecell[l]{\niia\ \\ (\kms)} & \makecell[l]{\Ox\ \\ (\kms)} & \makecell[l]{\OIII\ \\ (\kms)}\\
    \hline
    -1.2 & 355.0 & 273.0 & 263.0 & 301.0 & 216.0 & 302.0 \\
    1.2 & 230.0 & 201.0 & 199.0 & 203.0 & 206.0 & 268.0 \\
    2.0 & 206.0 & 175.0 & 171.0 & 180.0 & 183.0 & 225.0 \\
    2.8 & 182.0 & 163.0 & 161.0 & 168.0 & 168.0 &  \\
    3.5 & 176.0 & 155.0 & 154.0 & 158.0 & 164.0 &  \\
    4.5 & 171.0 & 155.0 & 149.0 & 151.0 & 159.0 &  \\
    5.9 & 147.0 & 137.0 & 133.0 & 141.0 & 120.0 &  \\
    \hline
    \end{tabular}
    \begin{tablenotes}[flushleft]
	\small {
	\item[] Notes: All values are obtained from fitting a single Gaussian peak to the line profile.
    }
    \vspace{0.1cm}
    \end{tablenotes}
    \hrule
\end{threeparttable}
\end{center}

\newpage
\section{Additional tables and figures}
\label{section:tables_figures}

\begin{figure}[h!]
\centering
\includegraphics[width=8.5cm, trim= 0cm 0cm 0cm 0cm, clip=true]{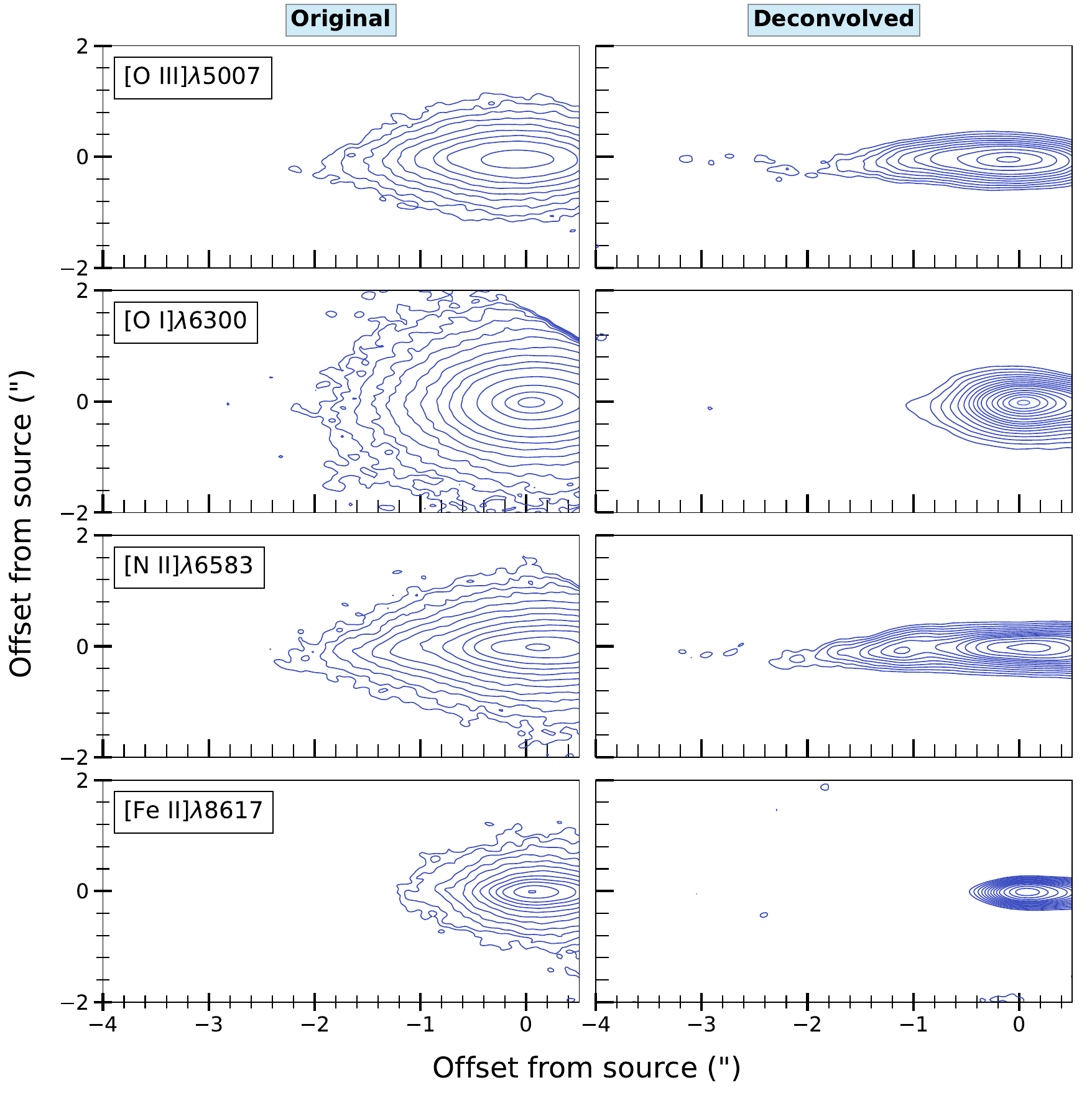}
\includegraphics[width=8.5cm, trim= 0cm 0cm 0cm 0cm, clip=true]{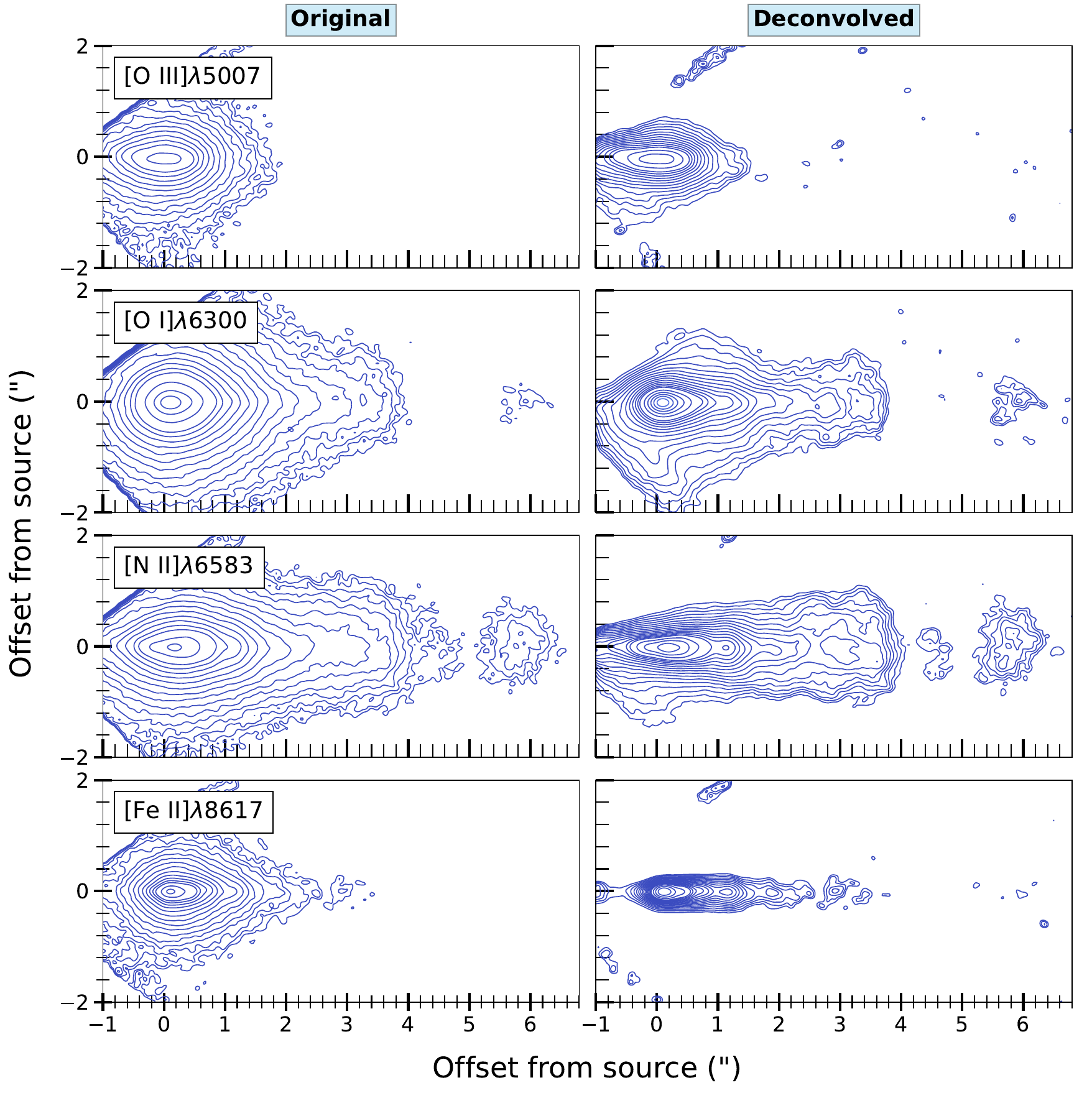}
\caption{\label{fig:decon_ims_app}Additional FEL spectro-images. Left: the blue-shifted jet channels (-140 \kms\ to +90 \kms) in Field B, before and after deconvolution. Contours are as in Fig. \ref{fig:decon_blue}, starting at 3-$\sigma$ of the background rms level pre-deconvolution, approximately 10 \bunit\ (26 \bunit\ for the \OIII\ spectro-images due to higher background noise). In \Ox\ the starting contour is 2-$\sigma$ (10 \bunit) to highlight the faint emission in this lobe. Right: the same for the red-shifted velocity channels (-90 \kms\ to +140 \kms) in Field A, with contours as in Fig. \ref{fig:decon_red}.}
\end{figure}

\begin{figure*}[h!]
\centering
\includegraphics[width=14cm, trim= 0.3cm 0.4cm 0.2cm 0cm, clip=true]{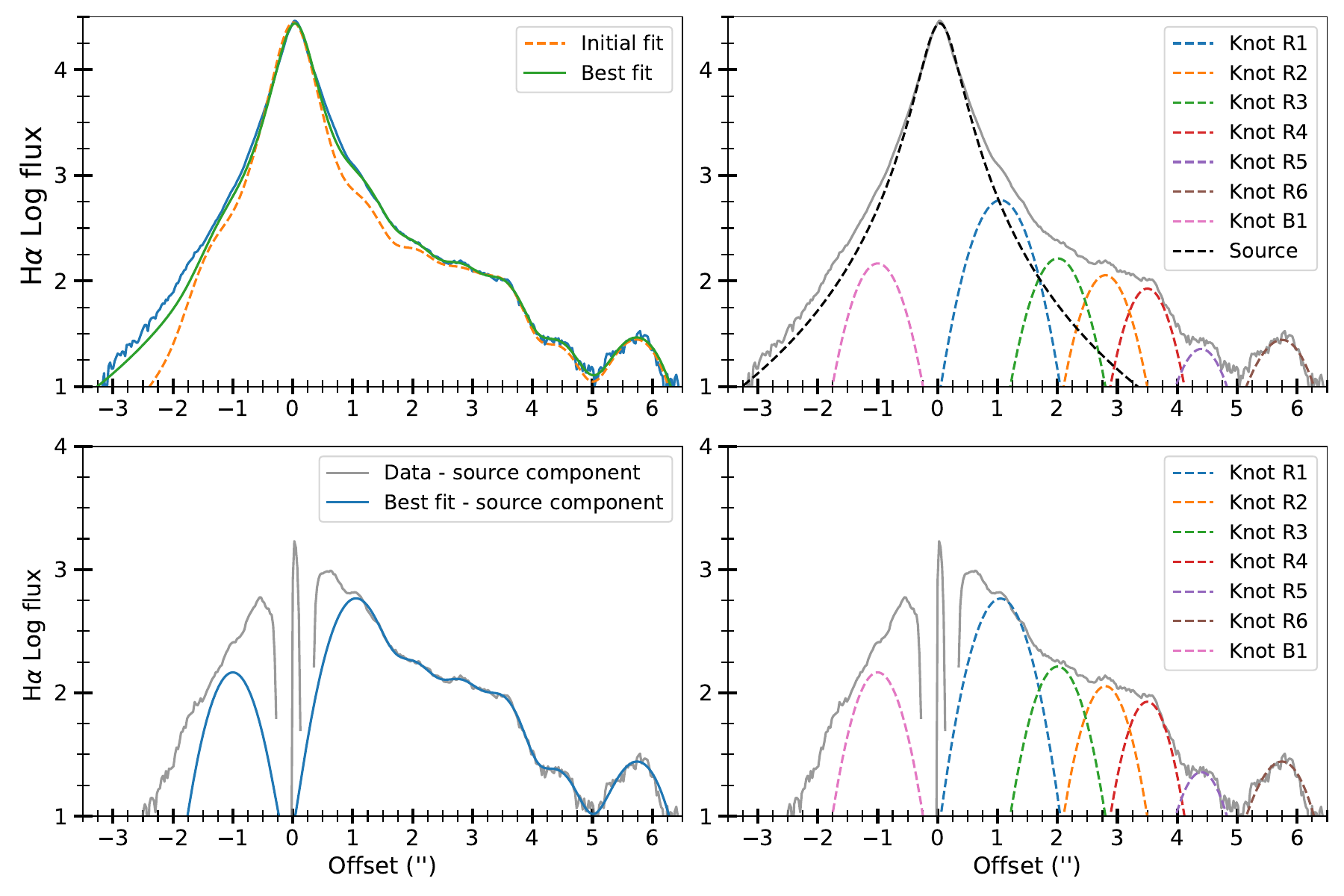}
\caption{\label{fig:flux_profile_models}Model fitting to the \Ha\ jet flux profile (grey) without scattered emission subtraction. The source peak is fitted with a Moffat function and knot peaks are fitted with Gaussian profiles. Top left: the initial and best-fit models compared with the data. Top right: the components of the best-fit model. Lower left: comparison of the best fit model and data after the fitted source peak is subtracted, highlighting the model residuals. Lower right: the data after the source peak is subtracted, overlaid with the remaining model components (the knot peaks).}
\end{figure*}

\begin{figure*}[h!]
\centering
\includegraphics[width=16cm, trim= 0.3cm 0.4cm 0.2cm 0cm, clip=true]{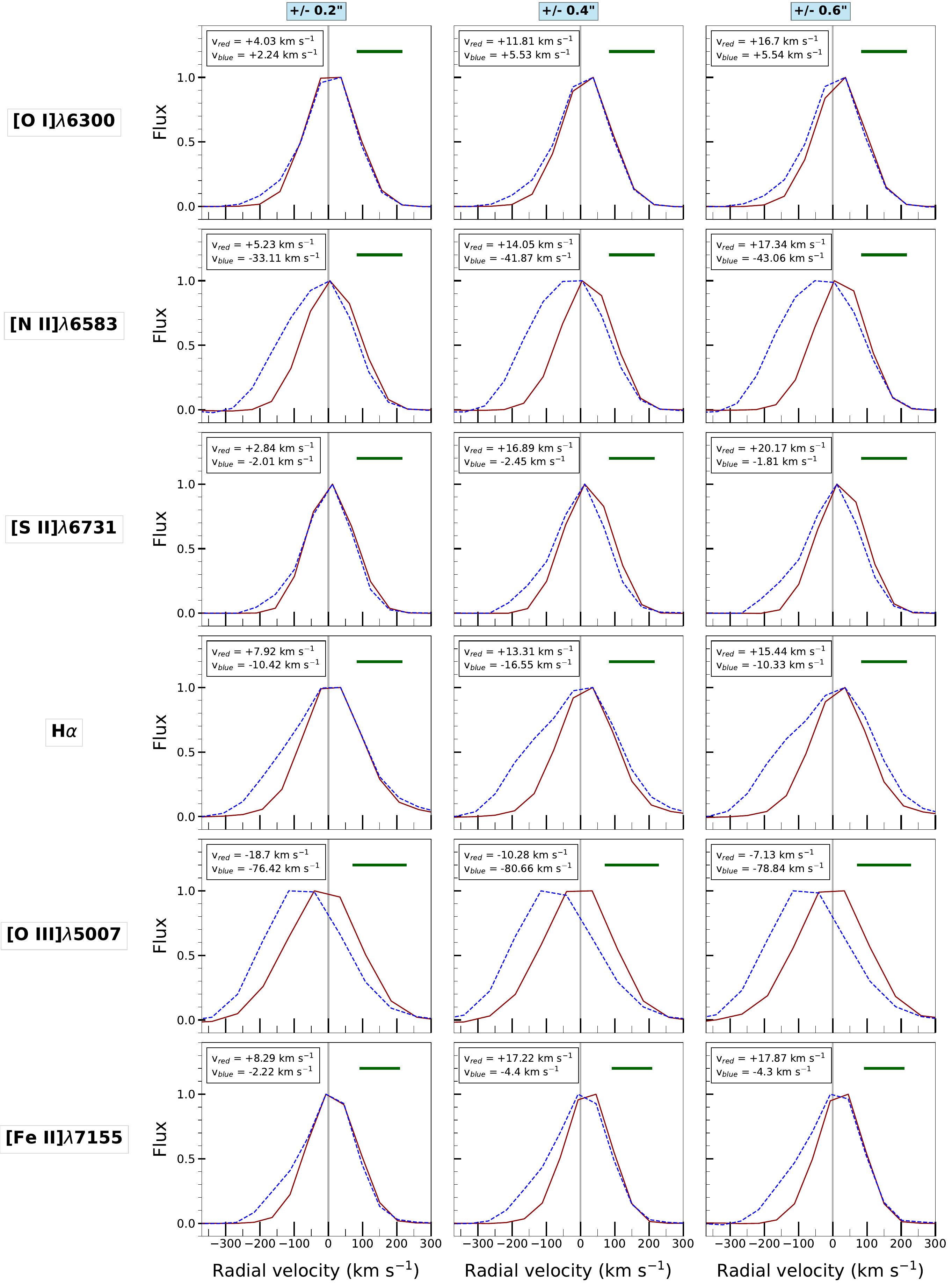}
\caption{\label{fig:profiles_0_2as}Emission line profiles sampled at offsets of +/- 0\farcs2, 0\farcs4 and 0\farcs6 in the red- and blue-shifted jet lobes (solid and dashed lines, respectively). Green bars indicate the approximate velocity resolution at the line wavelength.}
\end{figure*}
\newpage
\twocolumn

\begin{figure}[h!]
\centering
\includegraphics[width=8cm, trim= 0cm 0cm 0cm 0cm, clip=true]{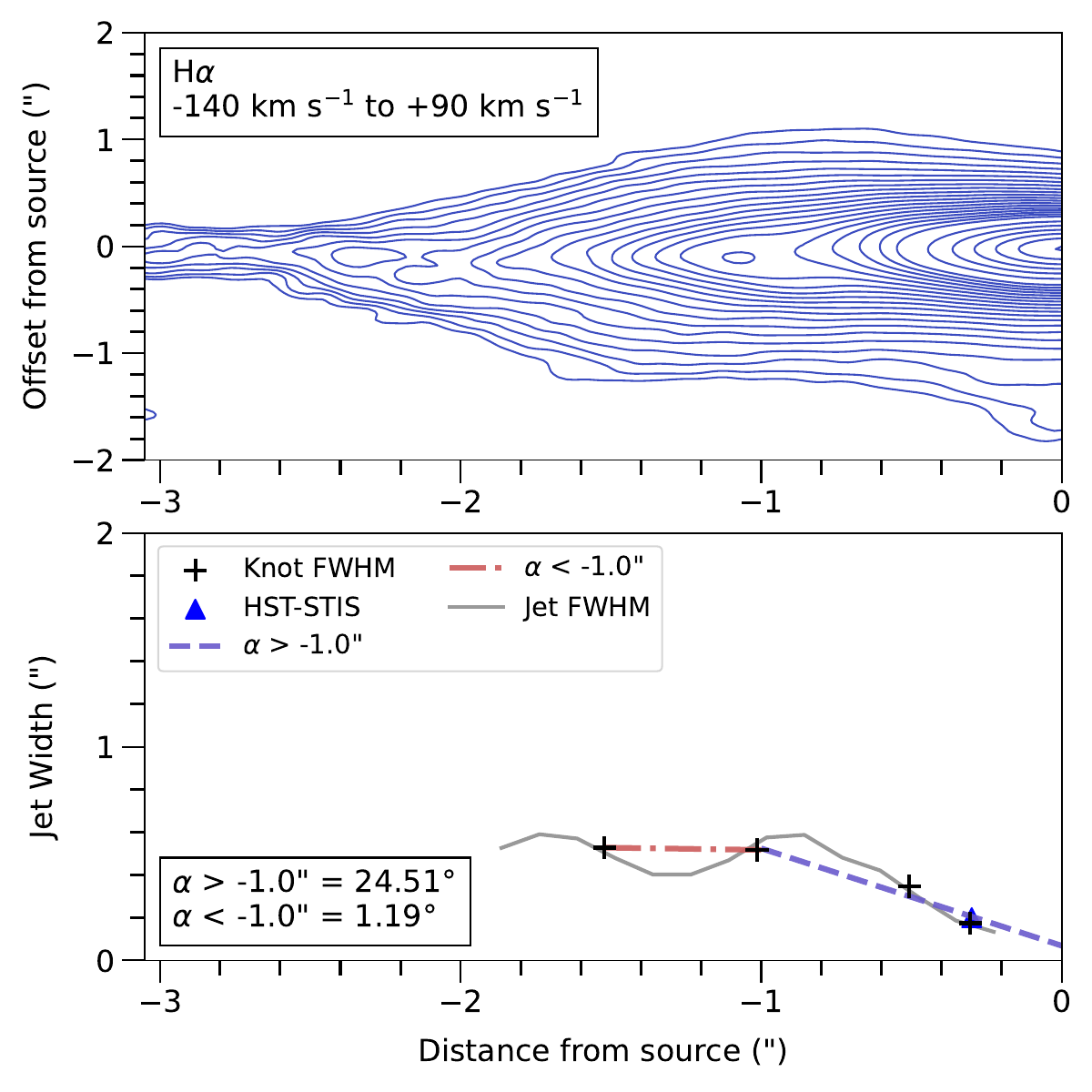}
\includegraphics[width=8cm, trim= 0cm 0cm 0cm 0cm, clip=true]{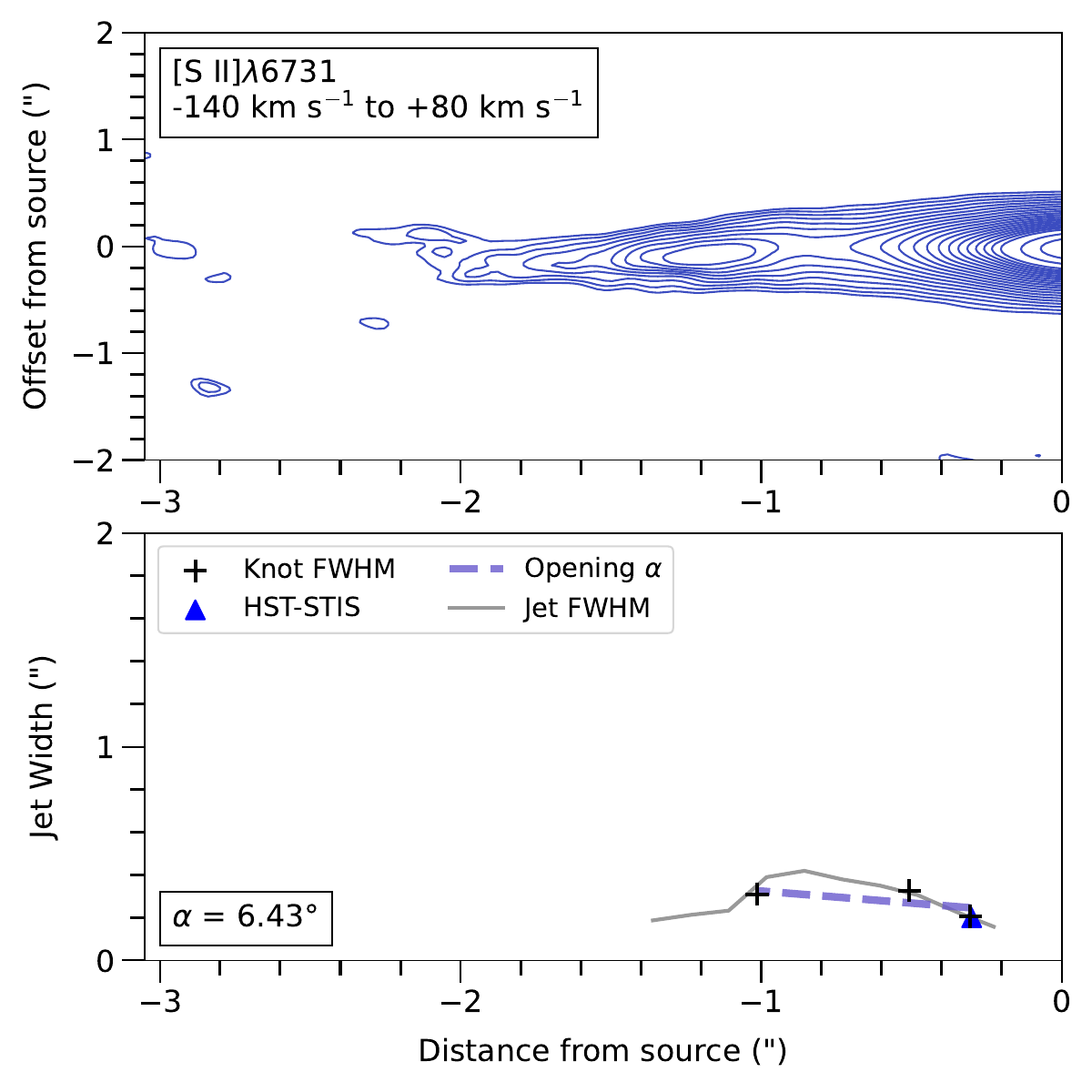}
\caption{\label{fig:bluejet_width_fitting}Deconvolved jet widths and opening angles $\alpha$ fitted across the full blue-shifted jet velocity range. Note that the fitted jet widths from HST data are also shown \citep{Coffey2004,Coffey2007}.}
\end{figure}

\begin{figure}
	\centering
	\includegraphics[width=8.5cm, trim= 0cm 0cm 0cm 0cm, clip=true]{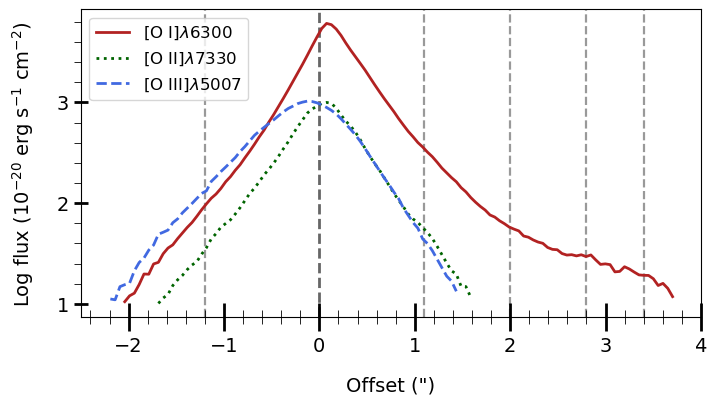}
	\caption{\label{fig:o_profiles} Flux profiles of the \Ox, [\oii]$\lambda$7330 and \OIII\ lines along the jet axis, without scattered emission subtraction. Vertical dashed lines indicate the source position (black) and the approximate knot positions (grey).}
\end{figure}

\begin{figure}
\centering

\includegraphics[width=8cm, trim= 0cm 0cm 0cm 0cm, clip=true]{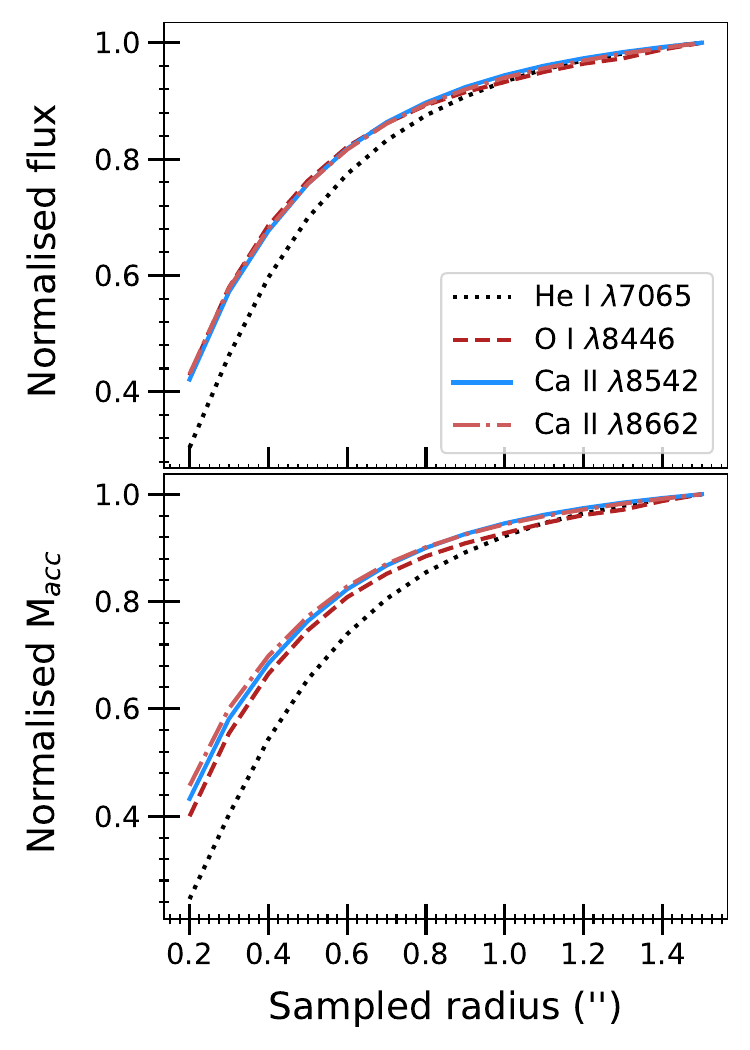}
\caption{\label{fig:macc_sampling}Comparison of emission line flux with increasing radius of the sampled region (top) and corresponding values of \Macc\ (bottom) for several accretion-tracing lines. }
\end{figure}

\begin{figure}
\centering
\includegraphics[width=8cm, trim= 0.3cm 0.4cm 0.2cm 0cm, clip=true]{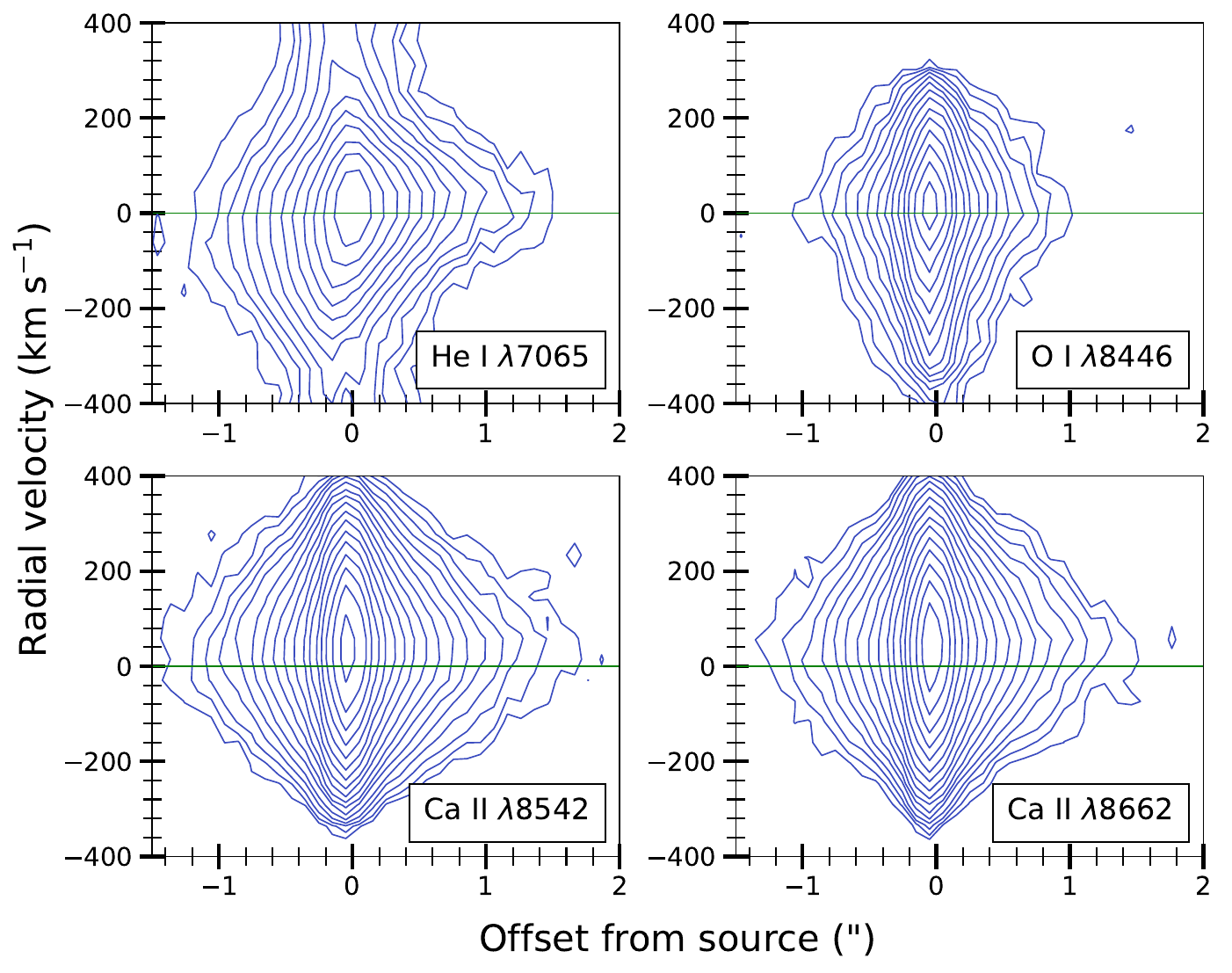}
\caption{\label{fig:macc_pvmaps}Position-velocity maps for the accretion-tracing lines sampled in Fig. \ref{fig:macc_sampling}. Contours start at 3$\sigma$ of the background level (4 $\times$ 10$^{-19}$ \ergscm) and scale as factors of $\sqrt{2}$.}
\end{figure}

\begin{figure}
	\centering
	\includegraphics[width=8cm, trim= 0cm 0cm 0cm 0cm, clip=true]{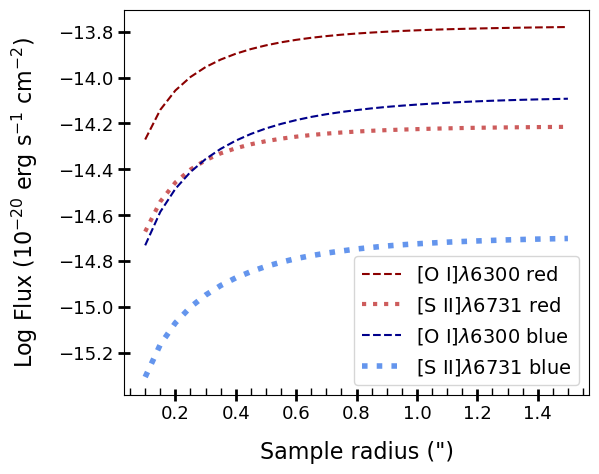}
	\caption{\label{fig:jet_flux_radii} FEL flux against sampling radius, for spectra centred at +/- 0\farcs5 in both the red- and blue-shifted lobes (denoted by red and blue lines, respectively).}
\end{figure}

\onecolumn

\begin{sidewaystable}
\begin{threeparttable}

\caption[Table of accretion luminosities and mass accretion rates]{Mass accretion rates obtained from the MUSE spectra in Sect. \ref{subsection:macc}, with \av\ taken to be 1.26 mag.}
\label{table:Macc_MUSE}
\renewcommand{\arraystretch}{1.6} 
\begin{tabular}{p{0.08\textwidth}<{\raggedright}   p{0.07\textwidth}<{\raggedright}   p{0.07\textwidth}<{\raggedright} 	  p{0.07\textwidth}<{\raggedright}  p{0.06\textwidth}<{\raggedright} 	p{0.05\textwidth}<{\raggedright}   p{0.08\textwidth}<{\raggedright} p{0.09\textwidth}<{\raggedright} 	p{0.08\textwidth}<{\raggedright}   p{0.06\textwidth}<{\raggedright}  p{0.02\textwidth}<{\raggedright}  p{0.06\textwidth}<{\raggedright\arraybackslash}} 
\hline \hline
\makecell[l]{Line}     &  \makecell[l]{$\lambda$ \\ (\AA)} & \makecell[l]{A}  &\makecell[l]{B}  & \makecell[l]{L$_{line}$ \\ \tiny{(10$^{-5}$ \Lsun)}} & \makecell[l]{$\Delta$L$_{line}$ \\ -} & \makecell[l]{L$_{acc}$ \\ \tiny{(10$^{-3}$ \Lsun)}}  & \makecell[l]{$\Delta$L$_{acc}$ \\  -} & \makecell[l]{$\dot M_{acc}$ \\ \tiny{(10$^{-8}$ \Msun\ yr$^{-1}$)}} & \makecell[l]{$\Delta$ $\dot M_{acc}$ \\ - } & Obsc. \\ 
\hline
H$\beta$		   & 4861.33 & 1.14 \tiny{$\pm$} 0.04 & 2.59 \tiny{$\pm$ 0.16} & 22.0 & 1.1 & 26.4 & 6.38 & 43.9 & 10.6 & 28 \\
\hei\ $\lambda$4922 & 4921.93  & 0.97 \tiny{$\pm$ 0.04} &  3.08 \tiny{$\pm$ 0.24} & 0.86 & 0.04 & 14.6 & 3.79 & 24.3 & 6.32 & 42\\
{[\oi]} 6300 & 6300.304 &  1.33 \tiny{$\pm$ 0.14} & 5.37 \tiny{$\pm$ 0.62} & 25.6 & 1.28 & 3.92 $\times$ 10$^{3}$ & 0.93 $\times$ 10$^{3}$	& 19.1	& 4.55 & - \\
H$\alpha$ &  6562.8 &  1.13 \tiny{$\pm$ 0.05} &  1.74 \tiny{$\pm$ 0.19} & 74.5 & 3.72 & 16.0 & 3.83	& 26.7 & 6.38 & 39 \\
\hei\ $\lambda$6678 & 6678.2 &  1.25 \tiny{$\pm$ 0.06} &  4.7 \tiny{$\pm$ 0.33} & 0.30 & 0.015 & 6.32 & 1.59 & 10.5	& 2.65 & 73 \\
\hei\ $\lambda$7065 & 7065.2 & 1.18 \tiny{$\pm$ 0.05} &  4.47 \tiny{$\pm$ 0.29} & 0.67 & 0.03 & 23.3 & 5.86 & 38.8 & 9.76 & 30 \\
\oi $\lambda$8446 & 8446.36 &  1.08 \tiny{$\pm$ 0.12} &  3.46 \tiny{$\pm$ 0.62} & 0.8 & 0.04 & 9.04 & 2.3 & 15.0 & 3.83 & 57 \\
\caii\ $\lambda$8498 & 8498.0 &  0.99 \tiny{$\pm$ 0.05} &  2.6 \tiny{$\pm$ 0.29} & 2.34 & 0.12 & 10.4 & 2.64 & 17.3 & 4.39 & 52 \\
\caii\ 8542 & 8542.09 &  0.97 \tiny{$\pm$ 0.06} &  2.43 \tiny{$\pm$ 0.29} & 2.73 & 0.14 & 10.1 & 2.56 & 16.7 & 4.26 & 53 \\
\caii\ $\lambda$8662 & 8662.14 & 0.93 \tiny{$\pm$ 0.06} & 2.30 \tiny{$\pm$ 0.3} & 2.17 & 0.11 & 9.17 & 2.36 & 15.3 & 3.93 & 57 \\
\hline
\end{tabular}
\begin{tablenotes}
\item[] Notes: Emission line luminosities (L$_{line}$), accretion luminosities (\Lacc), mass accretion rates (\Macc) and corresponding uncertainties ($\Delta$L$_{line}$, $\Delta$L$_{acc}$, and $\Delta$M$_{acc}$, respectively) are listed. Mass accretion rates obtained from permitted lines are quoted after correction using the obscuration factors in the last column. The coefficients A and B are those obtained by \citet{Alcala2014}, except in the case of the \Ox\ line where we use updated coefficients for the high-velocity jet from \citet{Nisini2018}.
\end{tablenotes}
\end{threeparttable}
\end{sidewaystable}

\end{appendix}
\end{document}